\documentclass[11pt]{article}
\usepackage[USenglish]{babel}
\usepackage[useregional=numeric,showisoZ=false,showseconds=false]{datetime2}
\usepackage[ascii]{inputenc}  %

\usepackage{multirow}
\usepackage{eepic,epic}
\usepackage{epsfig}
\usepackage{graphicx}
\usepackage[table]{xcolor}
\usepackage{afterpage,longtable}

\usepackage[hidelinks]{hyperref}
\usepackage{endnotes}

\usepackage{ifthen}
\usepackage{mathptmx}
\usepackage{amssymb}
\usepackage{amsfonts}
\usepackage{pifont}
\usepackage{amsmath}
\usepackage{nicefrac}

\newcommand{\lognote}[1]{}



\newcommand{\OMIT}[1]{} %

\newenvironment{proofs}{\noindent{\bf Proof.}\hspace*{1em}}{\literalqed\bigskip}
\def\literalqed{{\ \nolinebreak\hfill\mbox{\qedblob\quad}}}

\newenvironment{proofsof}[1]{\noindent{\bf Proof #1.}\hspace*{1em}}{\literalqed\bigskip}

\newcommand{\condition}{ \mid }

\newcommand{\Bibkeyhack}[3]{} %

\usepackage{ifthen}
\newcommand{\hugeDebug}{false}

\ifthenelse{\equal{\hugeDebug}{false}}{

}
{

}

\ifthenelse{\equal{\hugeDebug}{false}}{
\pagestyle{plain}
}
{
\pagestyle{headings}
}

\makeatletter
\newcommand{\singlespacing}{\let\CS=
\@currsize\renewcommand{\baselinestretch}{1}\tiny\CS}
\newcommand{\singlespacingplus}{\let\CS=
\@currsize\renewcommand{\baselinestretch}{1.25}\tiny\CS}
\newcommand{\doublespacing}{\let\CS=
\@currsize\renewcommand{\baselinestretch}{1.75}\tiny\CS}
\newcommand{\extradoublespacing}{\let\CS=
\@currsize\renewcommand{\baselinestretch}{1.9}\tiny\CS}
\newcommand{\draftspacing}{\let\CS=
\@currsize\renewcommand{\baselinestretch}{2.0}\tiny\CS}
\newcommand{\hugedraftspacing}{\let\CS=
\@currsize\renewcommand{\baselinestretch}{2.4}\tiny\CS}
\newcommand{\niceonespacing}{\let\CS=\@currsize\renewcommand{\baselinestretch}{1.1}\tiny\CS}
\newcommand{\nicetwospacing}{\let\CS=\@currsize\renewcommand{\baselinestretch}{1.2}\tiny\CS}
\newcommand{\nicethreespacing}{\let\CS=\@currsize\renewcommand{\baselinestretch}{1.3}\tiny\CS}
\newcommand{\singlespacingplusplus}{\let\CS=\@currsize\renewcommand{\baselinestretch}{1.35}\tiny\CS}
\newcommand{\nicefourspacing}{\let\CS=\@currsize\renewcommand{\baselinestretch}{1.4}\tiny\CS}
\newcommand{\nicefivespacing}{\let\CS=\@currsize\renewcommand{\baselinestretch}{1.5}\tiny\CS}
\newcommand{\nicesixspacing}{\let\CS=\@currsize\renewcommand{\baselinestretch}{1.6}\tiny\CS}
\newcommand{\nicesevenspacing}{\let\CS=\@currsize\renewcommand{\baselinestretch}{1.7}\tiny\CS}
\newcommand{\niceeightspacing}{\let\CS=\@currsize\renewcommand{\baselinestretch}{1.8}\tiny\CS}
\newcommand{\niceninespacing}{\let\CS=\@currsize\renewcommand{\baselinestretch}{1.9}\tiny\CS}
\makeatother

\makeatletter \@beginparpenalty=10000 \makeatother

\mathcode`\0="0030      %
\mathcode`\1="0031
\mathcode`\2="0032
\mathcode`\3="0033
\mathcode`\4="0034
\mathcode`\5="0035
\mathcode`\6="0036
\mathcode`\7="0037
\mathcode`\8="0038
\mathcode`\9="0039

\newcount\hour  \newcount\minutes  \hour=\time  \divide\hour by 60
\minutes=\hour  \multiply\minutes by -60  \advance\minutes by \time
\def\mmmddyyyy{\ifcase\month\or Jan\or Feb\or Mar\or Apr\or May\or Jun\or Jul\or
  Aug\or Sep\or Oct\or Nov\or Dec\fi \space\number\day, \number\year}
\def\hhmm{\ifnum\hour<10 0\fi\number\hour :%
  \ifnum\minutes<10 0\fi\number\minutes}
\makeatletter
\def\@cite#1#2{[#1\if@tempswa , #2\fi]}
\makeatother

\makeatletter
\def\@citex[#1]#2{\if@filesw\immediate\write\@auxout{\string\citation{#2}}\fi
  \def\@citea{}\@cite{\@for\@citeb:=#2\do
    {\@citea\def\@citea{,\linebreak[0]}\@ifundefined
      {b@\@citeb}{{\bf ?}\@warning
      {Citation `\@citeb' on page \thepage \space undefined}}%
\hbox{\csname b@\@citeb\endcsname}}}{#1}}
\makeatother

\makeatletter
\def\@cite#1#2{[#1\if@tempswa , #2\fi]}
\def\@citex[#1]#2{\if@filesw\immediate\write\@auxout{\string\citation{#2}}\fi
  \def\@citea{}\@cite{\@for\@citeb:=#2\do
    {\@citea\def\@citea{,\kern1pt\linebreak[0]}\@ifundefined
      {b@\@citeb}{{\bf ?}\@warning
      {Citation `\@citeb' on page \thepage \space undefined}}%
\hbox{\csname b@\@citeb\endcsname}}}{#1}}
\makeatother

\newcommand\qedblob{\mbox{\ding{113}}}

\newcommand{\score}{\mathit{score}}

\newtheorem{theorem}{Theorem}[section]

\newtheorem{corollary}[theorem]{Corollary}
\newtheorem{claim}[theorem]{Claim}
\newtheorem{corollaryhacked}[theorem]{Corollary (to the proofs 
of Thms./Cor.~3.2/3.6/3.9/3.12/3.14/3.16--3.18 
of~\protect\cite{car-cha-hem-nar-tal-wel:j:sct})}
\newtheorem{definition}[theorem]{Definition}

\newtheorem{proposition}[theorem]{Proposition}

\newtheorem{construction}[theorem]{Construction}

\newcommand{\p}{\ensuremath{\mathrm{P}}}
\newcommand{\np}{\ensuremath{\mathrm{NP}}}
\newcommand{\conp}{\ensuremath{\mathrm{coNP}}}

\newcommand{\fp}{\ensuremath{\mathrm{FP}}}
\newcommand{\pf}{\ensuremath{\mathrm{PF}}}
\newcommand{\fnp}{\ensuremath{\mathrm{FNP}}}

\newcommand{\sigmastar}{\ensuremath{\Sigma^{\ast}}}

\usepackage[inline]{enumitem}

\newcommand{\generalsearchreduce}{\leq_{\textnormal{search}}}
\newcommand{\polysearchreduce}{\generalsearchreduce^{p}}
\newcommand{\searchreduce}{\generalsearchreduce^{p, \cale}}

\newcommand{\winners}{\textrm{Winners}}
\newcommand{\uniquewinner}{\textrm{UniqueWinnerIfAny}}

\newcommand{\sat}{\ensuremath{{\rm SAT}}}

\newcommand{\cale}{\ensuremath{\mathcal E}}

\newcommand{\calc}{\ensuremath{\mathcal C}}

\newcommand{\calt}{\ensuremath{\mathcal T}}

\newcommand{\caltone}{\calt_1}
\newcommand{\calttwo}{\calt_2}
\newcommand{\caltthree}{\calt_3}

\newcommand{\cc}{\ensuremath{\mathrm{CC}}}
\newcommand{\dc}{\ensuremath{\mathrm{DC}}}
\newcommand{\uw}{\ensuremath{\mathrm{UW}}}
\newcommand{\nuw}{\ensuremath{\mathrm{NUW}}}
\newcommand{\pv}{\ensuremath{\mathrm{PV}}}
\newcommand{\pc}{\ensuremath{\mathrm{PC}}}
\newcommand{\rpc}{\ensuremath{\mathrm{RPC}}}
\newcommand{\te}{\ensuremath{\mathrm{TE}}}
\newcommand{\tp}{\ensuremath{\mathrm{TP}}}

\newcommand{\ccpvtenuw}{\cc\hbox{-}\pv\hbox{-}\allowbreak\te\hbox{-}\allowbreak\nuw}

\newcommand{\ccpvtpnuw}{\cc\hbox{-}\pv\hbox{-}\allowbreak\tp\hbox{-}\allowbreak\nuw}

\newcommand{\ccpctenuw}{\cc\hbox{-}\pc\hbox{-}\allowbreak\te\hbox{-}\allowbreak\nuw}
\newcommand{\ccpctpuw}{\cc\hbox{-}\pc\hbox{-}\allowbreak\tp\hbox{-}\allowbreak\uw}
\newcommand{\ccpctpnuw}{\cc\hbox{-}\pc\hbox{-}\allowbreak\tp\hbox{-}\allowbreak\nuw}

\newcommand{\ccrpctenuw}{\cc\hbox{-}\rpc\hbox{-}\allowbreak\te\hbox{-}\allowbreak\nuw}
\newcommand{\ccrpctpuw}{\cc\hbox{-}\rpc\hbox{-}\allowbreak\tp\hbox{-}\allowbreak\uw}
\newcommand{\ccrpctpnuw}{\cc\hbox{-}\rpc\hbox{-}\allowbreak\tp\hbox{-}\allowbreak\nuw}

\newcommand{\dcpcteuw}{\dc\hbox{-}\pc\hbox{-}\allowbreak\te\hbox{-}\allowbreak\uw}
\newcommand{\dcpctenuw}{\dc\hbox{-}\pc\hbox{-}\allowbreak\te\hbox{-}\allowbreak\nuw}
\newcommand{\dcpctpuw}{\dc\hbox{-}\pc\hbox{-}\allowbreak\tp\hbox{-}\allowbreak\uw}
\newcommand{\dcpctpnuw}{\dc\hbox{-}\pc\hbox{-}\allowbreak\tp\hbox{-}\allowbreak\nuw}
\newcommand{\dcrpcteuw}{\dc\hbox{-}\rpc\hbox{-}\allowbreak\te\hbox{-}\allowbreak\uw}
\newcommand{\dcrpctenuw}{\dc\hbox{-}\rpc\hbox{-}\allowbreak\te\hbox{-}\allowbreak\nuw}
\newcommand{\dcrpctpuw}{\dc\hbox{-}\rpc\hbox{-}\allowbreak\tp\hbox{-}\allowbreak\uw}
\newcommand{\dcrpctpnuw}{\dc\hbox{-}\rpc\hbox{-}\allowbreak\tp\hbox{-}\allowbreak\nuw}

\newcommand{\caledash}{\ensuremath{\cale\hbox{-}}}
\newcommand{\pluralitydash}{\mathrm{Plurality}\hbox{-}}
\newcommand{\vetodash}{\mathrm{Veto}\hbox{-}}
\newcommand{\approvaldash}{\mathrm{Approval}\hbox{-}}

\newcommand{\pluralitydcpcteuw}{\pluralitydash\allowbreak\dc\hbox{-}\allowbreak\pc\hbox{-}\allowbreak\te\hbox{-}\allowbreak\uw}
\newcommand{\pluralitydcpctenuw}{\pluralitydash\allowbreak\dc\hbox{-}\allowbreak\pc\hbox{-}\allowbreak\te\hbox{-}\allowbreak\nuw}

\newcommand{\pluralitydcpctpnuw}{\pluralitydash\allowbreak\dc\hbox{-}\allowbreak\pc\hbox{-}\allowbreak\tp\hbox{-}\allowbreak\nuw}
\newcommand{\pluralitydcrpcteuw}{\pluralitydash\allowbreak\dc\hbox{-}\allowbreak\rpc\hbox{-}\allowbreak\te\hbox{-}\allowbreak\uw}
\newcommand{\pluralitydcrpctenuw}{\pluralitydash\allowbreak\dc\hbox{-}\allowbreak\rpc\hbox{-}\allowbreak\te\hbox{-}\allowbreak\nuw}

\newcommand{\pluralitydcrpctpnuw}{\pluralitydash\allowbreak\dc\hbox{-}\allowbreak\rpc\hbox{-}\allowbreak\tp\hbox{-}\allowbreak\nuw}

\newcommand{\vetodcpvteuw}{\vetodash\allowbreak\dc\hbox{-}\allowbreak\pv\hbox{-}\allowbreak\te\hbox{-}\allowbreak\uw}
\newcommand{\vetodcpvtenuw}{\vetodash\allowbreak\dc\hbox{-}\allowbreak\pv\hbox{-}\allowbreak\te\hbox{-}\allowbreak\nuw}

\newcommand{\vetodcpvtpnuw}{\vetodash\allowbreak\dc\hbox{-}\allowbreak\pv\hbox{-}\allowbreak\tp\hbox{-}\allowbreak\nuw}
\newcommand{\vetodcpcteuw}{\vetodash\allowbreak\dc\hbox{-}\allowbreak\pc\hbox{-}\allowbreak\te\hbox{-}\allowbreak\uw}
\newcommand{\vetodcpctenuw}{\vetodash\allowbreak\dc\hbox{-}\allowbreak\pc\hbox{-}\allowbreak\te\hbox{-}\allowbreak\nuw}

\newcommand{\vetodcpctpnuw}{\vetodash\allowbreak\dc\hbox{-}\allowbreak\pc\hbox{-}\allowbreak\tp\hbox{-}\allowbreak\nuw}
\newcommand{\vetodcrpcteuw}{\vetodash\allowbreak\dc\hbox{-}\allowbreak\rpc\hbox{-}\allowbreak\te\hbox{-}\allowbreak\uw}
\newcommand{\vetodcrpctenuw}{\vetodash\allowbreak\dc\hbox{-}\allowbreak\rpc\hbox{-}\allowbreak\te\hbox{-}\allowbreak\nuw}

\newcommand{\vetodcrpctpnuw}{\vetodash\allowbreak\dc\hbox{-}\allowbreak\rpc\hbox{-}\allowbreak\tp\hbox{-}\allowbreak\nuw}

\newcommand{\approvalccpcteuw}{\approvaldash\allowbreak\cc\hbox{-}\allowbreak\pc\hbox{-}\allowbreak\te\hbox{-}\allowbreak\uw}
\newcommand{\approvalccpctenuw}{\approvaldash\allowbreak\cc\hbox{-}\allowbreak\pc\hbox{-}\allowbreak\te\hbox{-}\allowbreak\nuw}
\newcommand{\approvalccpctpuw}{\approvaldash\allowbreak\cc\hbox{-}\allowbreak\pc\hbox{-}\allowbreak\tp\hbox{-}\allowbreak\uw}
\newcommand{\approvalccpctpnuw}{\approvaldash\allowbreak\cc\hbox{-}\allowbreak\pc\hbox{-}\allowbreak\tp\hbox{-}\allowbreak\nuw}
\newcommand{\approvalccrpcteuw}{\approvaldash\allowbreak\cc\hbox{-}\allowbreak\rpc\hbox{-}\allowbreak\te\hbox{-}\allowbreak\uw}
\newcommand{\approvalccrpctenuw}{\approvaldash\allowbreak\cc\hbox{-}\allowbreak\rpc\hbox{-}\allowbreak\te\hbox{-}\allowbreak\nuw}
\newcommand{\approvalccrpctpuw}{\approvaldash\allowbreak\cc\hbox{-}\allowbreak\rpc\hbox{-}\allowbreak\tp\hbox{-}\allowbreak\uw}
\newcommand{\approvalccrpctpnuw}{\approvaldash\allowbreak\cc\hbox{-}\allowbreak\rpc\hbox{-}\allowbreak\tp\hbox{-}\allowbreak\nuw}
\newcommand{\approvaldcpvteuw}{\approvaldash\allowbreak\dc\hbox{-}\allowbreak\pv\hbox{-}\allowbreak\te\hbox{-}\allowbreak\uw}
\newcommand{\approvaldcpvtenuw}{\approvaldash\allowbreak\dc\hbox{-}\allowbreak\pv\hbox{-}\allowbreak\te\hbox{-}\allowbreak\nuw}

\newcommand{\approvaldcpcteuw}{\approvaldash\allowbreak\dc\hbox{-}\allowbreak\pc\hbox{-}\allowbreak\te\hbox{-}\allowbreak\uw}
\newcommand{\approvaldcpctenuw}{\approvaldash\allowbreak\dc\hbox{-}\allowbreak\pc\hbox{-}\allowbreak\te\hbox{-}\allowbreak\nuw}
\newcommand{\approvaldcpctpuw}{\approvaldash\allowbreak\dc\hbox{-}\allowbreak\pc\hbox{-}\allowbreak\tp\hbox{-}\allowbreak\uw}
\newcommand{\approvaldcpctpnuw}{\approvaldash\allowbreak\dc\hbox{-}\allowbreak\pc\hbox{-}\allowbreak\tp\hbox{-}\allowbreak\nuw}
\newcommand{\approvaldcrpcteuw}{\approvaldash\allowbreak\dc\hbox{-}\allowbreak\rpc\hbox{-}\allowbreak\te\hbox{-}\allowbreak\uw}
\newcommand{\approvaldcrpctenuw}{\approvaldash\allowbreak\dc\hbox{-}\allowbreak\rpc\hbox{-}\allowbreak\te\hbox{-}\allowbreak\nuw}
\newcommand{\approvaldcrpctpuw}{\approvaldash\allowbreak\dc\hbox{-}\allowbreak\rpc\hbox{-}\allowbreak\tp\hbox{-}\allowbreak\uw}
\newcommand{\approvaldcrpctpnuw}{\approvaldash\allowbreak\dc\hbox{-}\allowbreak\rpc\hbox{-}\allowbreak\tp\hbox{-}\allowbreak\nuw}

\title{
Search Versus 
Search for Collapsing Electoral Control Types%
    }

\author{%
    Benjamin Carleton\thanks{Work done in part while at the University of Rochester's Department of Computer Science.}\\
    Department of Computer Science\\Cornell University\\Ithaca, NY 14850, USA
    \and
    Michael C. Chavrimootoo\footnotemark[1]~\thanks{Corresponding Author.}\\
    Department of Computer Science\\Denison University\\Granville, OH 43023, USA
    \and
    Lane A. Hemaspaandra\thanks{Work done in part while on a sabbatical visit to the University of D\"usseldorf.}\\
    Department of Computer Science\\University of Rochester\\Rochester, NY 14627, USA
    \and
    David E. Narv\'{a}ez\thanks{Work done in part while at the University of Rochester's Department of Computer Science and Virginia Tech's Bradley Department of Electrical and Computer Engineering.}\\
Faculty of Mathematics and Physics\\University of Ljubljana\\Ljubljana, Slovenia
    \and
    Conor Taliancich\footnotemark[1]\\
    Property Matrix\\
    Culver City, CA 90230, USA
    \and
    Henry B. Welles\\
    Department of Computer Science\\University of Rochester\\Rochester, NY 14627, USA
}

\date{July 7, 2022; revised November 11, 2025}%

\begin{document}\sloppy
\maketitle

\begin{abstract}
Electoral control types are ways of trying to 
change the outcome of elections by altering aspects of their 
composition and structure~\cite{bar-tov-tri:j:control}.
We say two compatible (i.e., having the 
same input types)
control types
that are about the same election system 
$\cale$ form a {collapsing pair} 
if for every possible input (which typically consists of a candidate set,
a vote set, 
a focus candidate, and sometimes other parameters related to
the nature of the attempted alteration), either both or neither of 
the attempted attacks can be successfully carried out (see the Preliminaries
for a more formal definition)~\cite{hem-hem-men:j:search-versus-decision}.
    For each of the seven general (i.e., holding for all election systems) electoral control type 
    collapsing pairs 
    found
    by Hemaspaandra, Hemaspaandra, and 
    Menton~\cite{hem-hem-men:j:search-versus-decision}
    and for each of the additional electoral control type 
    collapsing pairs
    of Carleton et al.~\cite{car-cha-hem-nar-tal-wel:j:sct}
    for veto
    and approval 
    (and many other election systems in light of that paper's Theorems~3.6 and~3.9),
   both members of the collapsing pair 
    have the same complexity since as sets they \emph{are} the same
    set. However, having the same complexity (as sets) is not enough to
    guarantee that as search problems they have the same complexity.
    In this paper, we explore the relationships between the \emph{search versions}
    of collapsing pairs. For each of the collapsing pairs of Hemaspaandra, 
    Hemaspaandra, and Menton~\cite{hem-hem-men:j:search-versus-decision} and Carleton et 
    al.~\cite{car-cha-hem-nar-tal-wel:j:sct}, we
    prove that the pair's members' 
    \emph{search-version}
    complexities are polynomially
    related (given access, for cases when the winner problem itself is not
    in polynomial time, to an oracle for the winner problem). Beyond that,
    we give efficient reductions that from a solution to one compute
    a solution to the other. 
   For the concrete
    systems plurality, veto, and
    approval, we
    completely determine
    which of their (due to our results) polynomially-related %
    collapsing search-problem pairs are 
    polynomial-time computable and which are
    NP-hard.        
\end{abstract}

\section{Introduction}

Algorithms for problems on elections are an 
important 
class
of combinatorial algorithms, 
in which 
an election's 
attacker
generally has available
an exponential number of 
potential 
actions.
Despite the combinatorially explosive number of 
potential 
actions 
available, 
in many 
cases polynomial-time algorithms 
can be developed.
Thus the study of manipulative actions against elections is a 
beautiful 
showcase of an area with real-world importance, where the 
outcome of the clash 
between finding efficient algorithms for combinatorially explosive problems, and proving the impossibility (if $\p\neq\np$) of finding such algorithms, 
is of vivid interest.

``Control'' attacks on elections try to make a focus candidate 
win/lose/uniquely-win/not-uniquely-win through such actions as
adding, deleting, or partitioning candidates or voters. Control was
first studied in the seminal work of Bartholdi, Tovey, and Trick~\cite{bar-tov-tri:j:control}, 
and has been further explored in many
papers (see the survey chapter by Faliszewski and Rothe~\cite{fal-rot:b:handbook-comsoc-control-and-bribery}).

Hemaspaandra, Hemaspaandra, and Menton~\cite{hem-hem-men:j:search-versus-decision} 
noted that, surprisingly, seven
pairs among the 44 (relatively) ``standard'' control types collapse:
For each election system (i.e., each mapping from candidates and votes to a winner
set among the candidates) $\cale$ whose vote type is linear orders,
for each of those seven pairs of types $\caltone$ and $\calttwo$
it holds that for each input there is a successful action on that input under the $\caltone$ control type
if and only if there is
a successful action on that input
under the $\calttwo$ control type. Viewed as sets, $\caltone$ and $\calttwo$
are identical!
Carleton et al.~\cite{car-cha-hem-nar-tal-wel:j:sct} noted that Hemaspaandra,
Hemaspaandra, and Menton's proof tacitly in fact established that those seven collapses
hold for \emph{all} election systems regardless of their vote type.

Carleton et al.~\cite{car-cha-hem-nar-tal-wel:j:sct} 
showed that those
seven pairs are the only collapsing pairs among the 44 standard control
types
if one wants the given collapse to hold for \emph{every} election system. 
However, 
that paper
discovered some additional 
collapses that hold specifically for veto or 
for approval voting, and also
found 
new
collapses that hold for all election systems that 
satisfy certain axiomatic properties.\footnote{%
Two of those axiomatic theorems had incorrect 
proofs in the 
Carleton et al.\ conference version~\cite{car-cha-hem-nar-tal-wel:c:sct} and/or 
its
associated \mbox{arXiv.org} full technical report~\cite{car-cha-hem-nar-tal-wel:t4-and-we-really-do-mean-to-cite-it-for-historical-accuracy:sct}, 
but the journal
version of 
that paper~\cite{car-cha-hem-nar-tal-wel:j:sct},
and the \mbox{arXiv.org} full technical report's May 2024 revisions,
change the axiomatic assumptions and then
correctly prove the thus modified results. 
The conference version was using the flawed
results to prove things about approval voting,
and fortunately the journal version, and the 
\mbox{arXiv.org} technical report's May 2024 revisions, show
that those consequences can still be 
obtained using the modified results.}

For each election system to which a collapse applies, the complexity of the
two collapsing control types is the same (since, due to collapsing, the two
control types yield the exact same set). 
However, Carleton et al.~\cite{car-cha-hem-nar-tal-wel:j:sct}
raised (even in its 
initial technical report version, which was 
the main motivation for the present paper)
as a challenge the issue of whether the \emph{search} complexity of the collapsing
pairs is also the same. That is the issue that is the focus of the 
present paper.

Why is it even plausible that sets with the same decision complexity
might have different search complexities (relative to some 
certificate/solution schemes)? Well, it indeed can and does happen if
$\p \neq \np \cap \conp$ (and so certainly happens if integer factorization
is not in polynomial time, since the natural decision version of that
is in $\np \cap \conp$).
In particular, if $\p \neq \np \cap \conp$, let $L$ be any fixed set in 
$(\np \cap \conp) - \p$. Suppose $L$'s finite alphabet is $\Sigma$. 
Consider a nondeterministic polynomial-time Turing machine (NPTM) $N$
over input alphabet $\Sigma$ that immediately halts and accepts. 
So $L(N) = \sigmastar$ and the certificate of acceptance (basically, the
accepting computation path) of $N$ on any input $x$ is the empty string.
Thus the function $s(x) = \epsilon$ solves this search problem. However,
consider an NPTM $N'$ over the alphabet $\Sigma$ that on each input $x$
nondeterministically simulates both the NPTM $N_L$ for $L$ and the 
NPTM $N_{\overline{L}}$ for $\overline{L}$. Clearly, $L(N') = \sigmastar$,
since each $x \in \sigmastar$ is either in $L$ or in $\overline{L}$.
However, the search complexity of $\sigmastar$ relative to the 
certificate scheme
of $N'$ is not polynomial, since
on input $x$ any accepting path of $N'$ will (after removing its initial 
guess bit regarding simulating $N_L$ or $N_{\overline{L}}$) yield an
accepting path for \emph{exactly} one of $N_L$ or $N_{\overline{L}}$
on input $x$, and so determines whether $x \in L$ or $x \not\in L$.
Thus, since $L \not\in \p$, no search function for $\sigmastar$ relative
to $N'$ can be polynomial-time computable. This argument is basically due
to a construction of Borodin and Demers~\cite{bor-dem:t:selfreducibility}
(see also~\cite{val:j:checking}, \cite{har-hem:j:up}, and the proof 
in~\cite[p.~39]{hem:j:juris-hartmanis-golden-rules-borodin-demers-proof} that (in an earlier appearance) is pointed to by the discussion in Footnote~9
of~\cite{hem-hem-men:j:search-versus-decision}).

This
paper explores the issue of whether the collapsing (as decision problems) pairs of
Hemaspaandra, Hemaspaandra, and Menton~\cite{hem-hem-men:j:search-versus-decision} 
and Carleton et al.~\cite{car-cha-hem-nar-tal-wel:j:sct} have the 
same \emph{search} complexity. We prove that in every case
the answer is Yes
(if given access, for the case where the election's winner problem
itself is not even in $\p$, to an oracle for the election's winner
problem), and indeed we show that in each case a solution for one can
(again, given access to the winner problem) 
be polynomial-time transformed
into a solution for the other. 
Thus the complexities are polynomially related (given access to the 
election system's winner problem).
Additionally, for the concrete cases of 
plurality's, veto's, and approval's collapsing pairs, we explore whether those 
polynomially equivalent 
search
complexities are clearly ``polynomial time,'' or are
$\np$-hard, and we resolve every such case.

Why is this important?  In reality, one typically---e.g., if one is a campaign manager or a corrupt vote collector/counter---wants 
not 
merely to efficiently compute whether 
there \emph{exists} some action that will make one's candidate win, but rather one wants to get one's hands, efficiently, on \emph{an actual such successful action}.  
Unfortunately, the literature's 
currently known 
results 
on collapsing 
control type pairs 
are each about
the ``exists'' case.
In contrast, this paper's results are establishing that the literature's existing collapsing 
control type pairs 
even
have the property that---given access to the winner problem for the election system in question\footnote{Although for the 
three 
concrete election systems we cover as examples that is not even needed as their winner problems are each in polynomial time.}---for both members of the pair the ``getting one's hands on a successful action when one exists'' issue is 
of the same complexity for both. We also show that one usually can 
efficiently use access to solutions for one to get solutions for the other.

\section{Preliminaries}
This section covers the preliminaries on election systems, control types/collapses, and search problems.

\subsection{Elections and Election Systems}\label{ss:elections}
An election consists of a finite candidate set $C$ and a finite collection of votes~$V$ over the
candidates in $C$. 
The ``type'' of the votes depends on the election system that one is considering.
Most typically, each vote is a linear ordering---a complete, transitive, asymmetric
binary relation---over the candidates (e.g., $3 > 1 > 2$). 
Another common vote type (called an approval vector) is that a vote is bit vector of 
length $\|C\|$, with each bit typically indicating approval~(1) or disapproval~(0) of
a candidate.
No two candidates can have the same name.\footnote{By allowing
candidates to have names, we 
are following the model of the papers
that this paper is most closely related 
to~\cite{hem-hem-men:j:search-versus-decision,car-cha-hem-nar-tal-wel:j:sct}.
It is possible that election systems in this model may exploit  candidate names in complicated ways.
However, adopting this model in fact
makes our results stronger than they would be if our model, for example, required election systems to be
\mbox{(candidate-)neutral} and/or to have 
the candidate identities in $(C,V)$ always be $1,2,\dots,\|C\|$.  
Also, studies on control are incompatible with 
assuming that candidate names are always 
$1,2,\dots,\|C\|$, since many control 
types, including all partition-based ones, 
change the candidate set, thus resulting in 
elections that do not satisfy that 
condition; that issue has been implicit ever since
the seminal work of
Bartholdi, Tovey, and Trick~\cite{bar-tov-tri:j:control}
that initiated the study of control attacks on elections.}

An election system maps from an election,
$(C, V)$, to a (possibly nonstrict) 
subset of $C$ (the set of winners). In pure social choice theory, empty winner sets are
usually excluded. However, in computational social choice theory 
empty winner sets are often
allowed, and in this paper we do allow %
empty winner sets.
In fact,
Bartholdi, Tovey, and Trick's~\cite{bar-tov-tri:j:control}
model of run-off partition of candidates, which is one focus
of this paper, would not even 
be well-defined on one-candidate elections if one viewed 
zero-winner elections as 
illegal.  That is, allowing empty winner
sets is compelled unless one wants to 
change---in ways that in fact would open 
other 
difficulties---definitions that have been broadly accepted and used 
for three decades.

Three particular election systems that we will discuss are plurality, veto, and approval elections.
In plurality elections, each vote is a linear ordering over $C$, and all candidates for whom the 
number of votes in which they were ranked first is the highest (possibly tied) among the candidates 
are winners.
In veto elections, each vote is a linear ordering over $C$, and all candidates for whom
the number of votes in which they were ranked last is the lowest (possibly tied) among the 
candidates are winners. 
In approval elections, each vote is an approval vector, and all candidates who garner
the most (possibly tied) approvals are the election's winners.

\subsection{Control Types and Collapses}

Definition~\ref{def:control-types} and the text following it define the 24 partition-based control types
(for each election system $\cale$). For uniformity,
we take the definition essentially verbatim from the papers we are most closely related 
to~\cite{hem-hem-men:j:search-versus-decision,car-cha-hem-nar-tal-wel:j:sct},
which themselves were drawing on the line of earlier papers---starting with Bartholdi, Tovey, 
and Trick~\cite{bar-tov-tri:j:control}---that developed the current set of control notions
(for more history and citations, 
see~\cite{fal-rot:b:handbook-comsoc-control-and-bribery,hem-hem-men:j:search-versus-decision,%
car-cha-hem-nar-tal-wel:j:sct}). One must be clear as to the handling of candidates who are 
tied winners in first-round elections. The two tie-handling rules typically studied as to 
partition-based control types are
ties-eliminate (\te)---in which a candidate must uniquely win its subelection to proceed to the 
next 
round---and ties-promote (\tp)---in which all winners of a subelection proceed to the next 
round.
In Definition~\ref{def:control-types} and later in the paper, we will at times 
speak of
elections
whose candidate set is $C'$ but whose votes, due 
for example to
candidate partitioning and/or first-round candidate eliminations, are over 
a set $C \supseteq C'$.
For example, 
for the $(C_1,V)$ of part~\ref{part:blort-2} of Definition~\ref{def:control-types},
the $C'$ is $C_1$ 
and the votes in $V$ are over $C$.
As is standard in the literature (e.g., this was used implicitly in the seminal
control paper~\cite{bar-tov-tri:j:control} and many other control papers since, see also~\cite[Footnote~4]{car-cha-hem-nar-tal-wel:j:sct} where it is explicit), in such cases we always take this to mean that the
votes are each masked down to just the candidates in $C'$.

\begin{definition}[see~\cite{hem-hem-men:j:search-versus-decision,car-cha-hem-nar-tal-wel:j:sct}
	and the references/history therein]\label{def:control-types}
Let $\cale$ be an election system.
\begin{enumerate}
    \item In the constructive control by partition of voters problem for $\cale$, in the \tp\ or \te\ tie-handling rule model
    (denoted by $\caledash\allowbreak\ccpvtpnuw$ or $\caledash\allowbreak\ccpvtenuw$, respectively), 
    we are given an election $(C, V)$, and a candidate $p \in C$. We ask if there is a 
    partition\footnote{%
    	A partition of a multiset $V$ is a pair of multisets $V_1$ and $V_2$ such that $V_1 \cup 
    	V_2 = V$, where $\cup$ denotes multiset union.
    	A partition of a set $C$ is a pair of sets $C_1$ and $C_2$ such that $C_1 \cup C_2 = C$ and
    	$C_1 \cap C_2 = \emptyset$, where $\cup$ and $\cap$ are standard set union and 
    	intersection.}
    of $V$ into $V_1$ and $V_2$ such that $p$ is a winner of the two-stage election where the winners of
   subelection $(C, V_1)$ that survive the tie-handling rule compete 
   (with respect to vote collection $V$) 
    against the winners of subelection $(C, V_2)$ that survive the tie-handling rule.
    Each election (in both stages) is conducted using election system $\cale$.
    
    \item\label{part:blort-2} In the constructive control by run-off partition of candidates problem for $\cale$, in the \tp\ or \te\  tie-handling rule model
    (denoted by $\caledash\allowbreak\ccrpctpnuw$ or $\caledash\allowbreak\ccrpctenuw$, respectively), 
    we are given an election $(C, V)$, and a candidate $p \in C$. We ask if there is a 
    partition of $C$ into $C_1$ and $C_2$ such that $p$ is a winner of the two-stage
    election where the winners of
    subelection $(C_1, V)$ that survive the tie-handling
    rule compete 
    (with respect to vote collection $V$)
    against the winners of 
    subelection $(C_2, V)$
    that survive the tie-handling rule.
    Each election (in both stages) is conducted using election system $\cale$.
    
    \item In the constructive control by partition of candidates problem for $\cale$, in the \tp\ or \te\  tie-handling rule model
    (denoted by $\caledash\allowbreak\ccpctpnuw$ or $\caledash\allowbreak\ccpctenuw$, respectively), 
    we are given an election $(C, V)$, and a candidate $p \in C$. We ask if there is a 
    partition of $C$ into $C_1$ and $C_2$ such that $p$ is a winner of the two-stage
    election where the winners of
    subelection $(C_1, V)$ that survive the tie-handling
    rule compete (with respect to vote collection $V$) against 
    all candidates in $C_2$.
    Each election (in both stages) is conducted using election system $\cale$.
    \end{enumerate}
\end{definition}

In each of the six control types defined above, we can
replace ``is a winner'' with ``is a unique winner'' to denote
constructive control in the unique-winner (\uw) model. The resulting
control types are appended with ``-\uw'' rather than ``-\nuw.'' $\nuw$ denotes
the so-called nonunique-winner model, aka, the cowinner model, in which merely being 
an overall winner, whether tied or not, is the goal. 
This completes the definition of the
12 standard constructive partition-control types
(for each election system; in text discussions, we will often omit the ``for each
election system,'' and so may speak of control of a given type generically).
The standard 12 additional destructive partition-control types differ only in that instead of 
seeking to make a focus candidate a winner or unique winner, the goal is to \emph{prevent} the focus
candidate from being a winner or unique winner. To denote those, we replace the ``\cc'' with
a  ``\dc,''
e.g., the destructive control by
partition of voters problem for $\cale$
using the TE handling rule and the UW model is denoted $\cale$-DC-PV-TE-UW\@.
Thus the 24 standard types of partition-based control are defined. As is standard,
each control type is quietly defining---in some sense \emph{is}---a set, namely, the set of all
inputs on which the given control attack can be successfully carried out. Thus, for example,
$\vetodcpvtpnuw$ is a set. We will immediately use this view in the next sentence.

For any
two control types $\caltone$ and $\calttwo$ that are compatible (i.e., have the same collection of input fields), Carleton et al.~\cite{car-cha-hem-nar-tal-wel:j:sct}
say that 
$\caltone$ and $\calttwo$ collapse if $\caltone = \calttwo$.
If $\caltone$ and $\calttwo$ are both about the same election system $\cale$ (i.e., each has as its 
prefix ``$\cale$-'') and $\caltone$ and $\calttwo$ are compatible, we will say that control types
$\caltone$ and $\calttwo$ are $\cale$-matched.
If control types $\caltone$ and $\calttwo$ are both about 
the same 
election system $\cale$ 
and 
collapse (and so also are compatible), we say that they are a 
\emph{collapsing pair}.
Since all 24 control types just defined are mutually compatible, and none of the other $44-24=20$ 
standard control types are known to be involved in \emph{any} of the collapses found in
Hemaspaandra, Hemaspaandra, and Menton~\cite{hem-hem-men:j:search-versus-decision}
and Carleton et al.~\cite{car-cha-hem-nar-tal-wel:j:sct}, we will not further mention 
compatibility in this paper.
(See~\cite{car-cha-hem-nar-tal-wel:j:sct} for the standard definitions of the 20 control types not defined here.)

Before giving the collapses found by 
Hemaspaandra, Hemaspaandra, and Menton~\cite{hem-hem-men:j:search-versus-decision}
and 
Carleton et al.~\cite{car-cha-hem-nar-tal-wel:j:sct}, we give the definitions of three axiomatic properties of election systems. 
An election system $\cale$ is said to satisfy Property~$\alpha$ (see~\cite{che:j:rational-selection,sen:j:choice-functions}) if for every election $(C, V)$ and every 
candidate $p\in C$
it holds that  if $p$ is a winner of election $(C,V)$ then
for every $C'$
satisfying $p\in C'\subseteq C$ candidate $p$ remains a winner in the restriction of $(C,V)$ to $C'$.
An election system $\cale$ is said to satisfy Property Unique-$\alpha$, the unique version of 
Property~$\alpha$,
if for every election $(C, V)$ and every 
candidate $p\in C$
it holds that  if $p$ uniquely wins election $(C,V)$ then
for every $C'$
satisfying $p\in C'\subseteq C$ candidate $p$ 
uniquely wins
in the restriction of $(C,V)$ to $C'$.\footnote{%
What we call Property Unique-$\alpha$ (and for brevity we often drop the ``Property'') has been previously referred to as the ``Unique-WARP'' property. We discuss below
our reasons for moving away from that name.

Bartholdi et al.~\cite{bar-tov-tri:j:control} first used
the term WARP 
(Weak Axiom of Revealed Preferences) 
in the 
domain of computational social choice
by stating that it required that $p$ being a winner of election $(C, V)$ 
implies that $p$ remains 
a winner of every election $(C', V)$, where $p \in C' \subseteq C$, and they stated that ``this'' was also known as Property~$\alpha$. 
Unfortunately, Hemaspaandra et al.~\cite{hem-hem-rot:j:destructive-control} 
 apparently read that as saying 
that a
definition of WARP was being given. And so they
used
the term ``Unique-WARP'' when they defined the variant of 
that 
in which $p$ is required to be a unique winner.
In fact, WARP and Property~$\alpha$ are 
not equivalent.
(Sen~\cite{sen:j:choice-functions}, however, proved that 
WARP---in its long-settled and standard sense, see~\cite[Section~5.1]{han-gru:b:preferences-and-warp-in-stanford-encyclopedia}---is, with respect to the standard notion of 
choice functions (election systems) used in pure social choice theory,
equivalent to the combination of 
Property~$\alpha$ and a so-called Property~$\beta$ (from~\cite{sen:j:rational-choice}) that we will
not define here.  The reason we mention that the equivalence
holds for the pure social choice notion of election systems
is that in our model, where empty winner sets are allowed,
the equivalence in fact fails.
Consider the following election system $\cale$. If three or more candidates run, then the two candidates with the lexicographically smallest names win. Otherwise, no one wins.
Clearly, this election system does not satisfy Property~$\alpha$ (and certainly not Properties~$\alpha$ and~$\beta$ combined), but it does satisfy WARP\@.)

To summarize briefly, a
misreading in 
\cite{hem-hem-rot:j:destructive-control} of an 
ambiguous sentence in 
\cite{bar-tov-tri:j:control} led to an
infelicitous naming in~\cite{hem-hem-rot:j:destructive-control}
that we feel should be abandoned.
We thus use ``Unique-$\alpha$'' to denote 
what 
to date has been called 
``Unique-WARP'' 
(e.g., in~\cite{hem-hem-rot:j:destructive-control,erd:thesis:elections,erd-now-rot:j:sp-av,erd-fel-rot-sch:j:control-in-bucklin-and-fallback-voting,fit-hem-hem:j:control-manipulation,car-cha-hem-nar-tal-wel:c:sct}).
We do that so the terminology is analogous to what has long been used 
in social choice for the notion that inspired 
the Unique-$\alpha$ notion.%
}
Finally, an election system $\cale$ is said to be strongly voiced exactly if for every election $(C, V)$ with $C \neq \emptyset$, there is at least one winner of $(C, V)$ under $\cale$~\cite{full:hem-hem-rot:j:destructive-control}.

Here are the collapses found in 
Hemaspaandra, Hemaspaandra, and Menton~\cite{hem-hem-men:j:search-versus-decision}
and Carleton et al.~\cite{car-cha-hem-nar-tal-wel:j:sct}.
For every election system $\cale$, 
$\caledash\allowbreak\dcrpctpnuw=\caledash\allowbreak\dcpctpnuw$ and 
$\caledash\allowbreak\dcrpctenuw=\caledash\allowbreak\dcpctenuw=\caledash\allowbreak\dcrpcteuw=\caledash\allowbreak\dcpcteuw$ 
(\cite{hem-hem-men:j:search-versus-decision} supplemented by~\cite{car-cha-hem-nar-tal-wel:j:sct}'s
observation that the~\cite{hem-hem-men:j:search-versus-decision} collapse proofs work not just over
linear orders but in fact work over any election system regardless of its vote type).
For every election system $\cale$ satisfying Property Unique-$\alpha$, $\caledash\allowbreak\dcpcteuw=\caledash\allowbreak\dcrpctpuw=\caledash\allowbreak\dcpctpuw$~\cite{car-cha-hem-nar-tal-wel:j:sct}. And for every election $\cale$ that is strongly voiced and satisfies Property~$\alpha$, 
$\caledash\allowbreak\ccpctpuw=\caledash\allowbreak\ccrpctpuw$~\cite{car-cha-hem-nar-tal-wel:j:sct}.
Since approval voting is clearly strongly voiced and clearly satisfies Property~$\alpha$ and Property Unique-$\alpha$, these
four additional pair-collapses hold for approval voting~\cite{car-cha-hem-nar-tal-wel:j:sct}.
Note that $\approvaldcpcteuw$ participates in both the four-type and the three-type above collapses with
$\cale={\rm Approval}$, so by transitivity we have 
 the six-type collapse 
$\approvaldcrpctenuw=\approvaldcpctenuw = \approvaldcrpcteuw =$ $\approvaldcpcteuw = \approvaldcrpctpuw= \approvaldcpctpuw$~\cite{car-cha-hem-nar-tal-wel:j:sct}.

The remaining 
collapses from Carleton et al.~\cite{car-cha-hem-nar-tal-wel:j:sct} are
$\vetodcpvtenuw=\vetodcpvteuw$, $\approvaldcpvtenuw=\approvaldcpvteuw$, $\approvalccpctenuw = 
\approvalccrpctenuw$, $\approvalccpcteuw = \approvalccrpcteuw$, and 
finally,
$\approvalccpctpnuw=\allowbreak\approvalccrpctpnuw$.

\subsection{Search Problems and Their Interreductions and Complexity}

For each election system $\cale$, define the winner problem, $W_\cale$, by
$W_\cale = \{(C', V', p') \condition p' \in C'$ and $p'$ is a winner of the $\cale$ election $(C', 
V')\}$.
Our control types are each defined as a language problem---a set. However, each has a clear
``$\np^{W_\cale}$ search problem'' associated with it (so an ``$\np$ search problem'' if $\cale$'s  winner
problem is in $\p$). Namely, the problem of, on input $(C, V, p)$, outputting a partition
$(C_1, C_2)$ of the candidate set (or if the type is a voter partition type, a partition
$(V_1, V_2)$ of the vote 
collection) such that under that partition, the control action regarding $(C, V, p)$ is
successful under election system $\cale$.

To be able to speak clearly of search problems and their interrelationships and complexity, it is important
to be specific as to what we mean, both in terms of relating the solutions of two search problems and as to
classifying the complexity of a search problem. These two different tasks are related, yet differ in how they
are formalized.

Megiddo and Papadimitriou~\cite{meg-pap:tfnp} give a formalization of search problems along with
a notion of reductions between them. We will use that for connecting solutions of search problems
in this paper, except we will see that our problems will be connected in a way substantially tighter
than their framework anticipated. On the other hand, as we want to interconnect even electoral-control
search problems whose winner problems $W_\cale$ are potentially not in $\p$, we will not
require our relations to be polynomial-time decidable. 
(Rather, we will in effect require them to be
polynomial-time decidable given a $W_\cale$ oracle.
When $W_\cale \in \p$, that requirement of course is the same as requiring them to
be polynomial-time decidable.)

Following Megiddo and Papadimitriou~\cite{meg-pap:tfnp}, let $\Sigma$ be a finite alphabet with at least
two symbols and suppose $R \subseteq \sigmastar \times \sigmastar$ is a relation, decidable in polynomial
time relative to oracle $A$, that is polynomially-balanced. 
(Note: A relation is polynomially-balanced if there is a polynomial $p$ such that 
$(x, y) \in R$ implies $|y| \leq p(|x|)$.) That relation $R$ (which is decidable in $\p^A$) defines what 
Megiddo and Papadimitriou call the computational problem $\Pi_R$: Given an $x \in \sigmastar$, find some
$y \in \sigmastar$ such that $(x, y) \in R$ if such a $y$ exists, and reply ``no'' otherwise.
We will call the class of all such problems $\fnp^A$ (\cite{meg-pap:tfnp} in fact covered only the case
$A = \emptyset$, so their
class was called $\fnp$).

In the case of our 24 electoral-control problems, the ``$x$'' here is the problem input, $(C, V, p)$.  And 
if $(C, V, p)$ is a yes instance of the (viewed as a set) control type,
an
acceptable output $y$ in the sense of the above definition 
is (the encoding of) any partition $(C_1, C_2)$ of $C$ (or if the type is a voter partition, then any partition
$(V_1, V_2)$ of $V$) that leads to success under the control type, 
the input $(C,V,p)$, and 
the election system.
(In this paper we will not worry about encoding details, since such details are not a key issue here. We,
as is typical, merely assume reasonable, natural encodings.)

For each election system $\cale$ and each $\calt$ that is one of our 24 partition control types
for $\cale$, let $R_{\calt}$ be the natural $\p^{W_\cale}$-decidable, polynomially-balanced relation for its search
problem (i.e., $R_\calt$ is the set of pairs $((C, V, p), (P_1, P_2))$ where $(P_1, P_2)$ is a partition
of the problem's sort that in the setting $(C, V, p)$, under election system $\cale$, succeeds for the given
type of control action). Then
we will sometimes write $\Pi_{\calt}$ as a shorthand for $\Pi_{R_{\calt}}$. Note that for each of our 24
partition control types involving $\cale$, we clearly have $\Pi_{\calt} \in \fnp^{W_\cale}$, 
e.g., 
$\Pi_{\caledash\allowbreak\dcpctpnuw} \in \fnp^{W_\cale}$ holds (and,
for example, from that 
it follows immediately that, since $W_{\rm Approval} \in \p$,
$\Pi_{\approvaldcpctpnuw} \in \fnp$).

We wish to show that the known collapsing electoral control types also have polynomially related
(given access to an oracle for the winner problem for $\cale$) 
search complexity, and indeed
we wish to show
even that a solution for one can efficiently be used to generate a solution for the other. To do this,
we will need the notion of reductions between search problems. Fortunately, Megiddo and 
Papadimitriou defined a reduction notion between search problems that is close to what we need.
Megiddo and Papadimitriou~\cite{meg-pap:tfnp}
say that a reduction from problem $\Pi_R$ to problem $\Pi_S$ is a pair of polynomial-time
computable functions $f$ and $g$ such that, for any $x \in \sigmastar$, $(x, g(y)) \in R \iff (f(x), y) \in S$.
This in spirit is trying to say that we can map via $f$ to an instance $f(x)$ such that given a solution
relative to $S$ of $f(x)$ we can via $g$ map to a solution relative to $R$ of $x$.
Unfortunately, read as written, it does not seem to \emph{do} that, regardless of whether one
takes the omitted quantification over $y$ to be existential or to be universal. Either way the
definition leaves open a loophole in which on some $x$ for which there \emph{does} exist a solution
relative to $R$, the value of $f(x)$ will be some string that has no solution relative to $S$, and all
$g(y)$'s will be strings that are not solutions to $x$ relative to $R$. So the ``$\iff$'' will be satisfied since
both sides evaluate to False, but no solution transfer will have occurred. In the nightmare case, 
a given ``reduction'' could exploit this loophole 
on \emph{every} $x$ that has a solution relative to $R$.

In what follows, we will reformulate the definition more carefully, for our case, to close that loophole
(which is easily closed).
However, more interestingly, we will in effect alter the definition in two additional ways.
First, notice that in Megiddo and Papadimitriou's definition (assuming one fixes the above loophole first) the function
$f(x)$ allows the solution to $x$ on the $R$ side to be obtained via demanding a solution to a 
\emph{different} (than $x$) instance $f(x)$, on the $S$ side. But in our setting, collapsing types are the
\emph{same} set, just with differing ``witnessing'' relations. And our goal is to make connections via
those witnesses. So in the definitions we are about to give of our ``$\generalsearchreduce$'' reduction
family, we in effect require their $f(x)$ to be the identity function! Second, since we wish to connect
the solutions of collapsing pairs even when the winner problem of $\cale$ is not in $\p$, our reductions
will have the winner problem as an oracle. (The reason for this is that even given a $(C, V, p)$ and a 
partition $(C_1, C_2)$ of $C$ (or for voter partition types, a partition $(V_1, V_2)$ of $V$), to evaluate whether the partition
is a solution generally needs calls to $W_\cale$.)

We will not in these definitions explicitly write ``$\Pi$'' expressions, but the solutions we speak of
are with respect to the witnesses/actions that the given control type is about 
(they are the search component of $\Pi$, i.e., in our case they are those second components that appear in pairs, belonging to the underlying relation $R_\calt$, having ``$x$'' or ``$I$''
(i.e., the given $(C, V, p)$) as their first component), and so we view these definitions as a reformulated version
(with some changes for our particular situation) of the Megiddo and Papadimitriou ``$\Pi$'' approach.

Before we introduce our notation, let us first give some intuition behind it.
$\caltone \searchreduce \calttwo$ means $\calttwo$'s solutions are so powerful that
for problem instance $I$, given any solution for $\calttwo$ with respect to $I$ one can quickly, given
oracle $W_\cale$, build a solution to $\caltone$ with respect to $I$. That is why the notation has
$\caltone$ on the left.%

\begin{definition}\label{def:search-reduction}
    \begin{enumerate}
    \item\label{def:search-reduction-part-one}
    For an election system $\cale$ and $\cale$-matched, collapsing\footnote{The reason
    we include ``collapsing'' in the definition is that 
    if it is omitted, then one would trivially
    satisfy the notion on a given $I$ whenever
    $I$ was not in $\calttwo$.  That is, our
    notion is focused on pairs of types 
    that collapse---where each input is either 
    in both or 
    in neither.} %
    control types $\caltone$ and $\calttwo$, 
    we say that ``$\caltone$ is polynomially search-reducible to $\calttwo$ with respect to 
    $\cale$'' (denoted by $\caltone \searchreduce \calttwo$) if
    there is a reduction that, given an oracle for the winner problem for $\cale$, runs in
    polynomial time and on each input $(I, S)$, where $I$ is an input to $\caltone$
    and
    $S$ is a solution for $I$ with respect to $\calttwo$, outputs a solution
    $S'$ for $I$ with respect to $\caltone$.

    \item
    For an election system $\cale$ and $\cale$-matched, collapsing control types $\caltone$ and $\calttwo$,  we say that
    ``$\caltone$ is polynomially search-equivalent to $\calttwo$ with respect to $\cale$'' if
    $\caltone$ is polynomially search-reducible to $\calttwo$ with respect to $\cale$ and 
    $\calttwo$ is polynomially search-reducible to $\caltone$ with respect to $\cale$.
    \end{enumerate}
\end{definition}

\begin{definition}\label{def:poly-search-reduction}
    \begin{enumerate}
    \item\label{def:poly-search-reduction-part-one}
    For an election system $\cale$ and $\cale$-matched, collapsing control types $\caltone$ and $\calttwo$,
    we say that
    ``$\caltone$ is polynomially search-reducible to $\calttwo$'' 
    (denoted by $\caltone \polysearchreduce \calttwo$) if there is a reduction that runs in
    polynomial time and on each input $(I, S)$, where $I$ is an input to $\caltone$ and
    $S$ is a solution for $I$ with respect to $\calttwo$, outputs a solution
    $S'$ for $I$ with respect to $\caltone$.

    \item
    For an election system $\cale$ and $\cale$-matched, collapsing control types $\caltone$ and $\calttwo$, 
    we say that
    ``$\caltone$ is polynomially search-equivalent to $\calttwo$'' if
    $\caltone$ is polynomially search-reducible to $\calttwo$ and 
    $\calttwo$ is polynomially search-reducible to $\caltone$.
    \end{enumerate}
\end{definition}

The closest notion in the literature to our Definitions~\ref{def:search-reduction} and~\ref{def:poly-search-reduction} is the notion known as Levin reductions. 
Its definition can be found in such sources as Arora and Barak's 2009 textbook~\cite{aro-bar:b:complexity}, Piterman and Fisman's notes~\cite{fis-pit:url:notes-from-goldreich-1998} on a lecture from a 1998 course of Goldreich, and PlanetMath~\cite{hen:url:levin-reductions}. That notion 
much differs
from ours since it requires not just backward, but also forward transference of solutions.
However, Goldreich's 2008 textbook~\cite{gol:b:complexity} 
has a conflicting notion/definition of Levin reductions, and that notion
is close to our notion.
They are the same except 
we are focusing on collapsing electoral control
types and so sometimes make the winner problem available as an oracle, and our problem-to-problem reduction is simply the identity function.

When a polynomial-time reduction
has a polynomial-time computable
oracle, one can 
alter the reduction machine $M$'s action to have 
$M$ itself do without the oracle
by 
itself 
simulating the oracle's work.
We thus have the following 
observation.

\begin{proposition}\label{prop:skip-the-p-oracle}
    Let $\cale$ be an election system that has a polynomial-time winner problem,
    and let  $\caltone$ and $\calttwo$ be $\cale$-matched control types.
    \begin{enumerate}
    \item
    If $\caltone \searchreduce \calttwo$,
    then $\caltone \polysearchreduce \calttwo$.
   
    \item
     If $\caltone$ and $\calttwo$ are polynomially search-equivalent with
    respect to $\cale$, 
    then $\caltone$ and $\calttwo$ are polynomially search-equivalent.
    \end{enumerate}
\end{proposition}

The definitions we just gave provide the tools we will use to show that the complexities
of the two members of each known pair of collapsing
standard types are polynomially related to each other (given oracle access to $W_\cale$), 
and indeed that the members of the pair are closely related in terms of us being able to very
efficiently build a solution to 
one 
from a solution to the
other.\footnote{\label{f:altering-definitions}
Definitions~\ref{def:search-reduction}~part~\ref{def:search-reduction-part-one} 
and~\ref{def:poly-search-reduction}~part~\ref{def:poly-search-reduction-part-one} do not require that if
$S$ is not a solution to $\calttwo$ then the reduction declares that fact. Rather, the definitions are
simply about efficiently obtaining a solution to $\caltone$ given a solution to $\calttwo$. 
However, we mention that 
if one changed Definition~\ref{def:search-reduction}~part~\ref{def:search-reduction-part-one} to require 
detection of nonsolution-hood,
the set of pairs
$(\caltone, \calttwo)$ for which the reduction held would not change at all, since with the
$W_\cale$ oracle one can check whether $S$ is a solution to $\calttwo$.
Although Definition~\ref{def:poly-search-reduction}~part~\ref{def:poly-search-reduction-part-one} is not in general guaranteed to be unchanged if it is 
altered to require detection of the case where $S$ is not a solution to $\calttwo$, it clearly does remain
unchanged by that alteration whenever $W_\cale \in \p$. Most of the cases to which we apply 
Definition~\ref{def:poly-search-reduction} indeed satisfy $W_\cale \in  \p$; in
particular, plurality, veto, and approval voting each satisfy $W_\cale \in \p$.}

Note that even if $\cale$-matched collapsing types $\caltone$ and $\calttwo$
have polynomially related search complexities (in the above sense), that 
does not tell us whether they both are easy, or they both are hard.
Indeed, since for example our 
``close complexity relationship between the two members of the collapsing pair''
results regarding the seven Hemaspaandra, Hemaspaandra, and Menton~\cite{hem-hem-men:j:search-versus-decision}
collapses hold for \emph{all} election systems, it is
completely possible that though (as we will show)
the search complexity of such $\Pi_{\caltone}$ and $\Pi_{\calttwo}$ are  
polynomially related, for some election system $\cale'$ both could be easy 
and for some election system $\cale''$
both could be hard (indeed, even undecidable).

Nonetheless, at least for specific, concrete systems that collapse results exist for, it would be nice
to be able to prove results about whether their related search complexities are both
easy or are both hard. The $\Pi_{\calt}$ formalism,
even in the reformulated $\Pi$-free version in our definitions, is not ideal for doing this 
(since the Megiddo and Papadimitriou definition speaks of outputting ``no,'' which is not a typical
$\np$-machine action, and that definition also is speaking about various $y$ outputs without giving 
a framework for making clear how to speak of the instantiated (as to $y$) versions).
One could approach the issue in various ways, but we will do so by drawing on the flavor of some relations 
(sometimes called multivalued ``functions'') work that preceded that of Megiddo and
Papadimitriou~\cite{meg-pap:tfnp}. In particular, Book, Long, and Selman~\cite{boo-lon-sel:j:quant}
built a broad theory of multivalued $\np$ functions. Rather than presenting it here---it is a bit more
general than Megiddo and Papadimitriou~\cite{meg-pap:tfnp} in that it maps to path outputs rather than to 
certificates (certificates are in flavor closer to capturing a path's nondeterministic choices), and we 
do not need that generality---we draw just on one key notion. The Megiddo and Papadimitriou $\Pi_R$
model says that the task is to output ``any'' appropriate $y$. So in some sense, $\Pi_R$ is 
speaking of an entire family of
maps.
Viewed in the Book, Long, and Selman~\cite{boo-lon-sel:j:quant} lens, $\Pi_R$ is related to a multivalued function, call it $\widehat{\Pi}_R$,
that on a given input $x$ is undefined (i.e., maps to some special output $\perp$) if
$x$ has no solution $y$ (no appropriate-length $y$ with $(x, y) \in R$) 
and otherwise maps to the set of \emph{all} $y$ with 
$(x, y) \in R$. And, crucially, a ``single-valued refinement'' of that 
multivalued function is any function that on each 
input $x$:
\begin{enumerate}
    \item[(a)] is undefined on $x$ if $x$ has no solution relative to $R$, and
    \item[(b)] maps to exactly one solution of $x$ relative to $R$ if $x$ has at least one solution relative to $R$.
\end{enumerate}
For example, if $\Sigma=\{0,1\}$ and $R = \{(0x, 1) \condition x \in \sigmastar\} \cup
\{(0x, 0) \condition x \in \sigmastar\}$, then the refinements of $\widehat{\Pi}_R$ are each of the
uncountable number of functions that on inputs in $\{\epsilon\} \cup \{1x \condition x \in \sigmastar\}$
map to $\perp$ and on each input of the form $\{0x \condition x \in \sigmastar\}$ map to exactly one of
0 or 1.

We can now define what we mean by the search problem for $\calt$ to be easy or hard. We say that the
search problem for $\calt$ is polynomial-time computable if there \emph{exists} at least one refinement
of $\widehat{\Pi}_{\calt}$ (i.e., of $\widehat{\Pi}_{R_{\calt}}$) that is a polynomial-time computable function.
We say that the search problem for $\calt$ is $\np$-hard if for \emph{every} refinement $f$ of 
$\widehat{\Pi}_{\calt}$, it holds that $\np \subseteq \p^f$. We say that the search problem
for $\calt$ is $\np$-easy if there \emph{exists} at least one refinement $f$ of $\widehat{\Pi}_{\calt}$
that can be computed by a polynomial-time function given $\sat$ as its oracle,
i.e., is in the function class sometimes called ``$\fp^\np$'' or ``$\pf^{\np}$.'' If the search problem for $\calt$ is both $\np$-hard
and $\np$-easy, we say that the search problem is $\sat$-equivalent. 
Given two SAT-equivalent search problems, this definition does not promise (as, in contrast,
the second parts of Definition~\ref{def:poly-search-reduction} and, 
in some sense, 
Definition~\ref{def:search-reduction} do)
that one can produce a solution for one of the problems 
from any solution, for the same instance, to the other,
and vice versa.
However, the definitions just given do give us a clean way
to make clear that one of our search problems is hard, or is easy.

\section{Related Work}

The papers most related to this one are those discussed already: 
Hemaspaandra, Hemaspaandra, and Menton~\cite{hem-hem-men:j:search-versus-decision} 
and Carleton et al.~\cite{car-cha-hem-nar-tal-wel:j:sct} collapsed existing control types as decision problems.
And Carleton et al.~\cite{car-cha-hem-nar-tal-wel:j:sct}
posed as an open issue whether such
collapses also collapse the associated search problems as to their complexity. Addressing that issue
is the focus of the present paper.

Megiddo and Papadimitriou~\cite{meg-pap:tfnp} and Book, Long, and Selman~\cite{boo-lon-sel:j:quant}
set frameworks that we use and adapt, as to the study of search problems,
and refining multivalued functions.

Control was created by Bartholdi, Tovey, and Trick~\cite{bar-tov-tri:j:control}, and the 24 control types we
study were developed in that paper and (for the introduction of destructive control types and stating the \tp/\te\ rules)
in the work of Hemaspaandra, Hemaspaandra, and Rothe~\cite{hem-hem-rot:j:destructive-control}. The shift over time from
the earliest papers' focus on the \uw\ model to a focus on either both \uw\ and \nuw, or sometimes even
just \nuw, is discussed, with 
citations, by Carleton et al.~\cite[Related Work]{car-cha-hem-nar-tal-wel:j:sct}.

Many papers have studied control. Faliszewski and Rothe's survey~\cite{fal-rot:b:handbook-comsoc-control-and-bribery} is an excellent resource, with a rich range
of citations to papers investigating the control complexities of 
specific 
election systems.
Among the many systems that have been studied and seem to do somewhat well in resisting control attacks are
Schulze and ranked pair elections~\cite{par-xia:c:ranked-pairs,hem-lav-men:j:schulze-and-ranked-pairs}, 
Llull and Copeland elections~\cite{fal-hem-hem-rot:j:llull,fal-hem-hem:j:weighted-control}, 
normalized range voting~\cite{men:j:range-voting}, 
Bucklin and fallback elections~\cite{erd-rot:c:fallback,erd-pir-rot:c:open-problems} 
(see also~\cite{erd-fel-rot-sch:j:control-in-bucklin-and-fallback-voting}), and
SP-AV elections~\cite{erd-now-rot:j:sp-av}.
Election control, in many variants, continues to be an area of active interest~\cite{%
gup-roy-sau-zeh:j:resolute-control,%
yan:c:control-amendment-succ-winners,%
yan:c:two-stage-majoritarian-rules,%
col-gra-hid-mac-nav:c:divisiveness-in-rank-aggregation,%
kac-rot-tal:c:control-graph-restricted-wvg,%
kac-rot:control-without-changing-quotas,%
alo-ina-jai-tal-mor:c:control-liquid-democracy,%
kac-rot:c:weighted-voting-games,%
mau-nic-nus-rot-see:c:completing-picture-schulze-rp,%
yan:c:impact-your-paper,%
kac-rot:c:control-by-deleting-players-from-weighted-voting-games-is-np-pp-complete-for-the-penrose-banzhaf-power-index%
}.

Among the most popular election systems whose control complexity has been studied are 
approval~\cite{hem-hem-rot:j:destructive-control,bau-erd-hem-hem-rot:b:approval}, 
plurality~\cite{bar-tov-tri:j:control,hem-hem-rot:j:destructive-control}, 
and veto elections~\cite{lin:thesis:elections,mau-rot:j:control-veto-plurality,%
erd-reg-yan:c:puzzle}.
We will draw on some decision-problem algorithms from these papers to establish some of our 
search results. For example, Theorem~\ref{th:poly-time-computable-direct}
part~\ref{item:veto-collapse} follows directly from the properties of an algorithm of
Maushagen and Rothe~\cite{mau-rot:j:control-veto-plurality}.

\section{Results}\label{s:results}

This section will show that all known
collapsing control-type pairs even 
have the same \emph{search} complexity, 
given access to the election system's winner 
problem.  We will for the concrete cases
determine when 
that shared complexity level is ``polynomial-time computability,'' and when it is ``SAT-equivalence.'' 
And we also explore,
via our search-equivalence notions, how
solutions from one can be used to obtain
solutions to the other. We will even obtain new decision-case results (Proposition~\ref{p:np-complete}) that will help us in our quest to discover search complexities.

\subsection{Search Equivalences}

In the theorems and proofs that follow, $\winners_\cale(C, V)$ will denote the set of winners of the 
election $(C, V)$ under election system $\cale$.
$\uniquewinner_\cale(C, V)$ will be $\winners_\cale(C, V)$ if the latter has cardinality one, and 
will be the empty set otherwise.
To compute these sets, we will typically leverage access to (in settings where one has such access) the oracle,
$W_\cale = \{(C, V, p) \condition p \in C$ and $p$ is a winner of the $\cale$ election
$(C, V)\}$, since clearly each of the new functions can be computed in polynomial time given access to
$W_\cale$.

\begin{theorem}\label{th:hhm-small-class-equivalent}
    For every election system $\cale$, $\caledash\allowbreak\dcrpctpnuw$ and $\caledash\allowbreak\dcpctpnuw$ are 
    polynomially  search-equivalent with respect to $\cale$.
\end{theorem}
\begin{proofs}
    Let $\caltone = \caledash\allowbreak\dcpctpnuw$ and $\calttwo = \caledash\allowbreak\dcrpctpnuw$.
    
    $\caltone \searchreduce \calttwo$:
    We give a polynomial-time algorithm performing the reduction, given access to oracle $W_\cale$.
    On input $(I, S)$, where $I = (C, V, p)$ and $p \in C$, do the following. If the syntax is bad, halt.
    Otherwise, 
    check that $S$ is a 
    solution to $I$ under $\calttwo$. This may involve up to three calls to the oracle.
    If $S$ is not a good solution, then simply halt as the definition does not require anything in this case
    (although note also the comments in Footnote~\ref{f:altering-definitions}).
    Since $S$ is a solution and our model is \nuw, $p$ must have participated and lost either in
    one of the two first-round elections or (in fact, exclusive-or) in the final-round election. 
    In that election,
    let $C'$ be the candidate set (note that $p \in C'$ and $p$ is not a winner of $(C', V)$).
    Then output as the successful $\caltone$ solution
    $C_1 = C'$ and $C_2 = C-C'$. In $\caltone$ on input $I$, $p$ will be eliminated in the
    $(C', V)$ first-round election.\footnote{This case shows why it is important that our definition 
    is making
    an oracle to the winner problem of $\cale$ available. 
    Suppose we tried to claim that this direction held without any oracle use by,
    if $S$ is $(C_1, C_2)$, just
    outputting $S'=(C_1, C_2)$.
    But then if $p$ participated in and lost in the
    $(C_2, V)$ first-round $\calttwo$ case, $S'$ might not be a solution with respect to $\caltone$.
    Can we fix that by,
    if $p \in C_2$, just outputting $(C_2, C_1)$? No. Maybe in that case under $\calttwo$ candidate $p$
    lost in $(C_2, V)$, or maybe it lost in 
    $(\winners_\cale(C_2, V) \cup \winners_\cale(C_1, V), V)$. But if it was the latter, in
    $\caltone$ our second-round election is $(\winners_\cale(C_2, V)\cup C_1, V)$ and it is
    possible that $p$ wins in that.}
    
    $\calttwo \searchreduce \caltone$: We give the algorithm. On input $(I, S)$, where $I=(C, V, p)$
    and $S=(C_1, C_2)$, check for bad syntax and by using the $W_\cale$ oracle up to two times as per the
    nature of $\caltone$, check that $S$ is a solution to $I$ under $\caltone$. If the syntax is bad or $S$ is not a solution under $\caltone$, then simply halt.
    Otherwise,
    since $S$ is a solution under $\caltone$, $p$ was a participant in and eliminated either in the 
    $(C_1, V)$ contest or the $(C_2 \cup \winners_\cale(C_1, V), V)$ contest. In the former case,
    output $(C_1, C_2)$, and in the latter case let $D = C_2 \cup \winners_\cale(C_1, V)$ and output
    $(D, C-D)$. This is a solution for $\calttwo$.~\end{proofs}

\begin{proposition}\label{prop:search-reduce-transitive}
If $\cale$ is an election system, $\caltone$, $\calttwo$, and $\caltthree$ are pairwise $\cale$-matched
control types, $\caltone \searchreduce \calttwo$, and $\calttwo\searchreduce\caltthree$, then
$\caltone\searchreduce\caltthree$. That is, for $\cale$-matched types 
$\caltone$, $\calttwo$, and $\caltthree$, 
$\searchreduce$ is transitive.
\end{proposition}
\begin{proofs}
    Let $\cale$ be an election system and let $\caltone$, $\calttwo$, and $\caltthree$ be pairwise 
    $\cale$-matched
    control types such that $\caltone \searchreduce \calttwo$ via  $f$ and 
    $\calttwo\searchreduce\caltthree$ via $g$ (with $f$ and $g$ both running in polynomial time given oracle $W_\cale$). 
    We will show that 
    $\caltone \searchreduce\caltthree$. On input $(I, S)$, where $I=(C, V, p)$, if $S$ is a solution
    for $\caltthree$ on input $I$, then on input $((C, V, p) ,S)$ $g$ outputs a solution $S'$ for $\calttwo$ on input $I$.
    Additionally, if $S'$ is a solution
    for $\calttwo$ on input $I$, then on input $((C, V, p), S')$ $f$ outputs a solution $S''$ for $\caltone$ on input $I$.
    Thus if $S$ is a solution
    for $\caltthree$ on input $I$, then by applying $g$ and then $f$ in the manner just described we obtain, running in polynomial time
    with oracle $W_\cale$, a solution $S''$ for $\caltone$ on input 
    $I$. Thus $\caltone\searchreduce\caltthree$.~\end{proofs}

\begin{theorem}\label{th:hhm-large-class-equivalent}
For every election system $\cale$, and each pair $(\caltone, \calttwo)$ among the four control types
$\caledash\allowbreak\dcrpctenuw$, $\caledash\allowbreak\dcpctenuw$, $\caledash\allowbreak\dcrpcteuw$, and $\caledash\allowbreak\dcpcteuw$
(these decision-problem pairs are by~\cite{hem-hem-men:j:search-versus-decision} known to be collapsing),
we have
that $\caltone$ and $\calttwo$ are polynomially search-equivalent with respect to $\cale$.
\end{theorem}
\begin{proofs}
    We will make a closed cycle of $\searchreduce$ reductions involving these
    four types. In light of Proposition~\ref{prop:search-reduce-transitive}, this suffices to establish the theorem.  In each part, as per the reduction definition, we will assume our input is $(I, S)$, with
    $I = (C, V, p)$.
    
    $\caledash\allowbreak\dcrpcteuw \searchreduce \caledash\allowbreak\dcrpctenuw$:
    A solution $(C_1, C_2)$ for $I$ under $\caledash\allowbreak\dcrpctenuw$ is always a solution for $I$ under
    $\caledash\allowbreak\dcrpcteuw$, since \uw\ is stricter in the final round. If $p$ was eliminated under
    $\caledash\allowbreak\dcrpctenuw$, then $p$ is also eliminated under $\caledash\allowbreak\dcrpcteuw$.  So our 
    reduction here can simply output 
    the purported solution it is given.
    (Even if that is not a correct solution to the right-hand
    side, 
    the reduction's action is legal,
    since if given a nonsolution 
    as input all the reduction has to do, under the 
    definition of this reduction type,
    is not run for
    too long.)

    $\caledash\allowbreak\dcrpctenuw \searchreduce \caledash\allowbreak\dcpctenuw$:
    Say we are for $I=(C, V, p)$ given a purported solution $S = (C_1, C_2)$ to $\caledash\allowbreak\dcpctenuw$. 
     If $S$ is not a successful solution, immediately reject.
    Otherwise, either (a)~$p \in C_1$ but $p$ is not a unique winner of $(C_1, V)$, or
    (b)~$p \in \uniquewinner_\cale(C_1, V) \cup C_2$ yet $p$ is not a winner of 
    $(\uniquewinner_\cale(C_1, V) \cup C_2, V)$. Using our oracle, determine  which of 
    (a) or (b) holds (exactly one must hold if $S$ was a solution).
    If (a) holds, output $(C_1, C_2)$. This is then a successful solution on $I$ to $\caledash\allowbreak\dcrpctenuw$.
    If (b) holds, then set $D = \uniquewinner_\cale(C_1, V) \cup C_2$ and output $(D, C-D)$ and this
    is a successful solution of $I$ to $\caledash\allowbreak\dcrpctenuw$, since we know that $p$ is not a winner 
    of $(D, V)$, so it certainly is not a unique winner of 
    $(D,V)$,
    and so in our $\caledash\allowbreak\dcrpctenuw$
    first round $p$ participates and is eliminated.

    $\caledash\allowbreak\dcpctenuw \searchreduce \caledash\allowbreak\dcpcteuw$:
    Say we are given for $I=(C, V, p)$ a purported solution $S=(C_1, C_2)$ for $I$ to $\caledash\allowbreak\dcpcteuw$. 
    If $S$ is not a solution, immediately reject.
    Otherwise, we know that (a)~$p \in C_1$ and $p$ does not uniquely win in $(C_1, V)$, 
    exclusive-or
    (b)~$p \in \uniquewinner_\cale(C_1, V) \cup C_2$ yet $p$ is not a unique winner of 
    $(\uniquewinner_\cale(C_1, V) \cup C_2, V)$. If (a) holds, output $(C_1, C_2)$. This is a successful solution of $I$ for $\caledash\allowbreak\dcpctenuw$ as $p$ is eliminated in the first round. If
    (b) holds, set $D = \uniquewinner_\cale(C_1, V) \cup C_2$ and output $(D, C-D)$. This is a 
    successful solution of $I$ for $\caledash\allowbreak\dcpctenuw$ as $p$ will be eliminated
    in the first round.
    
    $\caledash\allowbreak\dcpcteuw \searchreduce \caledash\allowbreak\dcrpcteuw$:
    Say we are given $I=(C, V, p)$ and a purported solution $S=(C_1, C_2)$ for $\caledash\allowbreak\dcrpcteuw$ for
    $I$. If $S$ is not a solution, immediately reject.
    Otherwise, since $C_1$ and $C_2$ are symmetric in \rpc, w.l.o.g.\ assume $p\in C_1$
    (otherwise, if $p\in C_2$, rename $C_1$ and $C_2$ so that $p\in C_1$).
    So exactly one of (a) and (b) holds, where (a) and (b) are:
    (a)~$p \in C_1$ and $p$ is not a unique winner of $(C_1, V)$, 
    and
    (b)~$p\in \uniquewinner_\cale(C_1, V) \cup \uniquewinner_\cale(C_2, V)$ and 
    $p$ is not a unique winner of 
    $(\uniquewinner_\cale(C_1, V) \cup \uniquewinner_\cale(C_2, V), V)$.
    If (a) holds, output $(C_1, C_2)$ and that is a successful solution for $I$ of $\caledash\allowbreak\dcpcteuw$
    as $p$ 
    is eliminated in the first round. If (b) holds, set $D = \uniquewinner_\cale(C_1, V) \cup \uniquewinner_\cale(C_2, V)$, and output $(D, C-D)$ and that is a successful solution for $I$
    of $\caledash\allowbreak\dcpcteuw$ as $p$ 
    is eliminated in the first round.~\end{proofs}

In light of Proposition~\ref{prop:skip-the-p-oracle}, from Theorems~\ref{th:hhm-small-class-equivalent} and~\ref{th:hhm-large-class-equivalent}, we 
have, respectively, the two parts of the following corollary.

\begin{corollary}\label{cor:concrete-relationships-from-hhm}
For each $\cale \in \{{\rm Plurality},{\rm Veto},{\rm Approval}\}$ the following hold.
\begin{enumerate}
	\item $\caledash\allowbreak\dcrpctpnuw$ and $\caledash\allowbreak\dcpctpnuw$ are polynomially search-equivalent.
    \item For each pair $(\caltone, \calttwo)$ among the four types
    $\caledash\allowbreak\dcrpctenuw$, $\caledash\allowbreak\dcpctenuw$, $\caledash\allowbreak\dcrpcteuw$, and $\caledash\allowbreak\dcpcteuw$, 
    it holds that
    $\caltone$ and $\calttwo$ are polynomially 
    	search-equivalent.
\end{enumerate}
\end{corollary}

We now explore the system-specific collapses from Carleton et al.~\cite{car-cha-hem-nar-tal-wel:j:sct} in the same order 
that 
they appear
in that paper. We observe that the way they establish their new decision-problem collapses for veto, approval, and election systems
satisfying Unique-$\alpha$ is so constructive that the proofs implicitly show,
for each new collapsing decision-problem pair $(\caltone, \calttwo)$ that they find, that $\caltone$ and $\calttwo$ are polynomially search-equivalent, thus giving us 
the following corollary.

\begin{corollaryhacked}%
\label{cor:relationships-from-priors}~
\begin{enumerate}
    \item $\vetodcpvteuw$ and $\vetodcpvtenuw$ are polynomially search 
    equivalent.\label{item:veto-collapse}
    \item\label{cor:priors-part-two}
    For each election system $\cale$ that satisfies Unique-$\alpha$,
    $\caledash\allowbreak\dcpctpuw$ and $\caledash\allowbreak\dcpcteuw$ are polynomially
    search-equivalent.\footnote{
    One might expect here and in part~\ref{cor:priors-part-three} of this corollary the weaker
    conclusion ``polynomially search-equivalent with respect to $\cale$.'' But in both these parts we mean and prove ``polynomially search-equivalent.''}
    \item\label{cor:priors-part-three} For each strongly voiced election system $\cale$ that satisfies Property~$\alpha$,
    $\caledash\allowbreak\ccpctpuw$ and $\caledash\allowbreak\ccrpctpuw$ 
    are polynomially search-equivalent.
    \item\label{cor:priors-part-four}
    $\approvaldcrpctpuw$ and $\approvaldcpctpuw$ are polynomially search-equivalent.
    \item $\approvalccpctpnuw$ and $\approvalccrpctpnuw$ 
    are polynomially search-equivalent.
    \item $\approvaldcpvteuw$ and $\approvaldcpvtenuw$ are polynomially search-equivalent.
    \item $\approvalccpctenuw$ and $\approvalccrpctenuw$ are polynomially search-equivalent.
    \item $\approvalccpcteuw$ and $\approvalccrpcteuw$ are polynomially search-equivalent.
\end{enumerate}
\end{corollaryhacked}

\begin{proofs}
For all parts except~\ref{cor:priors-part-two} and~\ref{cor:priors-part-three}, we have
$W_\cale \in \p$ so we can check if the $S$ of the input is a valid solution, and so in those
parts we below assume that input always is a solution for $I=(C, V, p)$ of the problem
on the right-hand side of the reduction. For parts~\ref{cor:priors-part-two} and~\ref{cor:priors-part-three}, we cannot and do not make that assumption. 
\begin{enumerate}
    \item $\vetodcpvteuw \polysearchreduce \vetodcpvtenuw$: A solution $(C_1, C_2)$
    for $I$ under $\vetodcpvtenuw$ is always a solution for $I$ under
    $\vetodcpvteuw$, so we just output $(C_1, C_2)$.
    
    $\vetodcpvtenuw \polysearchreduce \vetodcpvteuw$: The proof of Theorem~3.2
    of~Carleton et al.~\cite{car-cha-hem-nar-tal-wel:j:sct} shows how to construct, given 
    $I=(C, V, p)$ and a solution $S=(C_1, C_2)$ for $\vetodcpvteuw$, a solution
    for $I$ to $\vetodcpvtenuw$, and we note that this construction can easily be done
    in polynomial time.

    \item Let $\cale$ be an election system that satisfies Unique-$\alpha$\@.
    The proof of Theorem~3.6 of 
    Carleton et al.~\cite{car-cha-hem-nar-tal-wel:j:sct} shows that
    $\caledash\allowbreak\dcpctpuw = \caledash\allowbreak\dcpcteuw = B_\cale = \{(C, V, p) \condition p \in C$ and $p$ is
    not a unique winner of the $\cale$ election $(C, V)\}$. They also show that for
    each $I \in B_\cale$, $(\emptyset, C)$ is a solution to $I$ for both
    $\caledash\allowbreak\dcpctpuw$ and $\caledash\allowbreak\dcpcteuw$.
    
    $\caledash\allowbreak\dcpctpuw \polysearchreduce \caledash\allowbreak\dcpcteuw$:
    Let our input be $(I,S)$. 
    If $S$ is a solution to $I$ for $\caledash\allowbreak\dcpcteuw$, then $I \in B_\cale$.
    So output $(\emptyset, C)$.
    (Note that there is no guarantee on the output if $S$ is not a solution of $I$
    for $\caledash\allowbreak\dcpcteuw$, and that is fine since our reduction type does not require us to make any such guarantee. This fact implicitly holds throughout the rest of this proof and so we do not mention it again.)

    $\caledash\allowbreak\dcpcteuw \polysearchreduce \caledash\allowbreak\dcpctpuw$: 
    Let our input be 
    $(I,S)$.
    If $S$ is a solution to $I$ for $\caledash\allowbreak\dcpctpuw$, then $I \in B_\cale$.
    So output $(\emptyset, C)$.

    \item Let $\cale$ be an election system that is strongly voiced and that satisfies Property~$\alpha$\@.
    The proof of Theorem~3.9 of
    Carleton et al.~\cite{car-cha-hem-nar-tal-wel:j:sct} shows that
    $\caledash\allowbreak\ccpctpuw = \caledash\allowbreak\ccrpctpuw = A_\cale = \{(C, V, p) \condition p \in C$ and $p$ is
    the unique winner of the $\cale$ election $(C, V)\}$.  They also show that for
    each $I \in A_\cale$, $(\emptyset, C)$ is a solution to $I$ for both
    $\caledash\allowbreak\ccpctpuw$ and $\caledash\allowbreak\ccrpctpuw$.
    
    $\caledash\allowbreak\ccpctpuw \polysearchreduce \caledash\allowbreak\ccrpctpuw$:
    Let our input be 
    $(I,S)$.
    If $S$ is a solution to $I$ for $\caledash\allowbreak\ccrpctpuw$, then $I \in A_\cale$.
    So output $(\emptyset, C)$.
    
    $\caledash\allowbreak\ccrpctpuw\polysearchreduce\caledash\allowbreak\ccpctpuw$:
    Let our input be 
    $(I,S)$.
    If $S$ is a solution to $I$ for $\caledash\allowbreak\ccpctpuw$, then $I \in A_\cale$.
    So output $(\emptyset, C)$.

    \item 
    The proof of Theorem~3.12 of
    Carleton et al.~\cite{car-cha-hem-nar-tal-wel:j:sct} shows that 
    $\approvaldcpctpuw = \approvaldcrpctpuw = B = \{(C, V, p) \condition p \in C$ and $p$ is not a unique winner of the approval election $(C, V)\}$. 
    They also show that for each $I \in B$, $(\emptyset, C)$ is a solution to $I$
    for both $\approvaldcpctpuw$ and $\approvaldcrpctpuw$.
    
    $\approvaldcpctpuw \polysearchreduce \approvaldcrpctpuw$:
    Let our input be 
    $(I,S)$.
    Since $S$ is a solution to $I$ for $\approvaldcrpctpuw$, $I \in B$.
    So output $(\emptyset, C)$.
    
    $\approvaldcrpctpuw\polysearchreduce\approvaldcpctpuw$:
    Let our input be 
    $(I,S)$.
    Since $S$ is a solution to $I$ for $\approvaldcpctpuw$, $I \in B$.
    So output $(\emptyset, C)$.

    \item
    The proof of Theorem~3.14 of
    Carleton et al.~\cite{car-cha-hem-nar-tal-wel:j:sct} shows that 
    $\approvalccpctpnuw = \approvalccrpctpnuw = A = \{(C, V, p) \condition p \in C$ and $p$ is 
    a winner
    of the approval election $(C, V)\}$. 
    They also show that for each $I \in A$, $(\emptyset, C)$ is a solution to $I$
    for both $\approvalccpctpnuw$ and $\approvalccrpctpnuw$.
    
    $\approvalccpctpnuw \polysearchreduce \approvalccrpctpnuw$: 
    Let our input be 
    $(I,S)$.
    Since $S$ is a solution to $I$ for $\approvalccrpctpnuw$, $I \in A$.
    So output $(\emptyset, C)$.
    
    $\approvalccrpctpnuw\polysearchreduce\approvalccpctpnuw$: 
    Let our input be 
    $(I,S)$.
    Since $S$ is a solution to $I$ for $\approvalccpctpnuw$, $I \in S$.
    So output $(\emptyset, C)$.

    \item $\approvaldcpvteuw\polysearchreduce\approvaldcpvtenuw$: On input
    $I=(C, V, p)$ and $S$, since a solution for $I$ to $\approvaldcpvtenuw$ is a solution
    for $I$ to $\approvaldcpvteuw$, output $S$.
    
    $\approvaldcpvtenuw\polysearchreduce\approvaldcpvteuw$: The proof of Theorem~3.16 of 
    Carleton et al.~\cite{car-cha-hem-nar-tal-wel:j:sct} shows how to construct, given $I=(C, V, p)$ and a solution $S$
    for $\approvaldcpvteuw$,
    a solution to $I$ for $\approvaldcpvtenuw$, and we note that the construction can be done in polynomial time.

    \item $\approvalccpctenuw\polysearchreduce\approvalccrpctenuw$:
    The proof of Theorem~3.17
    of Carleton et al.~\cite{car-cha-hem-nar-tal-wel:j:sct} shows how to construct, given $I=(C, V, p)$
    and a solution $S$ for $\approvalccrpctenuw$,
    a solution to $I$ for $\approvalccpctenuw$, and we note that the construction can be done in polynomial time.
    
    $\approvalccrpctenuw\polysearchreduce\approvalccpctenuw$:
    The proof of Theorem~3.17 of Carleton et al.~\cite{car-cha-hem-nar-tal-wel:j:sct} 
    shows how to construct, given $I=(C, V, p)$
    and a solution $S$ for $\approvalccpctenuw$,
    a solution to $I$ for $\approvalccrpctenuw$, and we note that the construction can be done in polynomial time.

    \item $\approvalccpcteuw\polysearchreduce\approvalccrpcteuw$:
    The proof of Corollary~3.18 of Carleton et al.~\cite{car-cha-hem-nar-tal-wel:j:sct} 
    shows how to construct, given $I=(C, V, p)$ and a solution $S$ for $\approvalccrpcteuw$,
    to construct a solution to $I$ for $\approvalccpcteuw$, and we note that the construction can be done in polynomial time.

    $\approvalccrpcteuw\polysearchreduce\approvalccpcteuw$:
    The proof of Corollary~3.18
    of Carleton et al.~\cite{car-cha-hem-nar-tal-wel:j:sct} shows 
    how to construct, given $I=(C, V, p)$ and solution $S$ for $\approvalccpcteuw$,
    to construct a solution to $I$ for $\approvalccrpcteuw$, and we note that the construction can be done in polynomial time.
\end{enumerate}~\end{proofs}%

\pagebreak[3]

\begin{corollary}\label{cor:from-uwarp-to-approval}
	\begin{enumerate}
		\item For each pair $(\caltone, \calttwo)$ among the six types
		$\approvaldcrpctenuw$, $\approvaldcrpcteuw$, $\approvaldcpctenuw$,
		$\approvaldcpcteuw$, $\approvaldcrpctpuw$, and $\approvaldcpctpuw$,
		it holds that $\caltone$ and $\calttwo$ are polynomially search-equivalent.
		\item $\approvalccpctpuw$ and $\approvalccrpctpuw$ are polynomially
		search-equivalent.
	\end{enumerate}
\end{corollary}
\begin{proofs}
    \begin{enumerate*}[itemjoin=\\]
		\item This follows directly from 
		Corollaries~\ref{cor:concrete-relationships-from-hhm} 
		and~\ref{cor:relationships-from-priors} 
		(Parts~\ref{cor:priors-part-two} and~\ref{cor:priors-part-four} of the latter)
		since approval satisfies Unique-$\alpha$.
		\item This follows directly from Corollary~\ref{cor:relationships-from-priors}
		since approval is strongly voiced and satisfies Property~$\alpha$.
	\end{enumerate*}~\end{proofs}

In the next subsection, for each concrete search-equivalence above (i.e., those about plurality, 
veto, and approval), we pinpoint the exact search 
complexity
of the associated search problems
by proving them to be either polynomial-time computable or SAT-equivalent
(see Table~\ref{table:summary}).

\subsection{Concrete Search Complexities}

In the proof of Theorem~\ref{th:approval-immune-poly-time} we will draw on the notion of, and known
results about, immunity. The notion dates back to the seminal work of Bartholdi, Tovey, and 
Trick~\cite{bar-tov-tri:j:control} (see also~\cite{hem-hem-rot:j:destructive-control},
which introduced the destructive cases and fixed a minor flaw in the original formulation for the
unique-winner case). We take the definitions' statements essentially verbatim from
Carleton et al.~\cite{car-cha-hem-nar-tal-wel:j:sct}.
In the unique-winner model, we say an election system is immune to a particular type of control
if the given type of control can never 
change a candidate from not uniquely winning to uniquely winning
(if the control type is constructive) or change a candidate from uniquely winning to not uniquely winning (if the control type is destructive).
In the nonunique-winner model, we say an election system is immune to a particular type of control
if the given type of control can never
change a nonwinner to a winner (if the control type is constructive) or change a winner to a nonwinner (if the control type is destructive).

\begin{theorem}\label{th:approval-immune-poly-time}
For each $\calt \in \{\approvaldcpctpnuw, \approvaldcpcteuw,\allowbreak \approvalccpctpuw, \approvalccpctpnuw\}$, 
it holds that the search problem for $\calt$ is polynomial-time computable.
\end{theorem}
\begin{proofs}
    Consider the case of $\approvaldcpcteuw$. Approval is known to be immune to $\dcpcteuw$~\cite{hem-hem-rot:j:destructive-control}. So on each input $(C, V, p)$, we have:
    If $p$ is a unique winner of election $(C, V)$ under approval voting, then $p \not\in \approvaldcpcteuw$ 
    (i.e., there exists no candidate partition under which $p$ is not a final-round unique winner in the $\pc\hbox{-}\te$ two-stage election process under approval voting). Our polynomial-time search-problem algorithm thus is the following:
    On input $(C, V, p)$, in polynomial time determine whether $p$ is a unique winner under approval of election
    $(C, V)$. If it is, output $\perp$ (indicating there is no partition that will prevent $p$ from being the unique winner in the $\pc\hbox{-}\te$ two-stage election process under approval, regarding $(C, V)$). Otherwise output as our
    solution the partition $(\emptyset, C)$, since this will make the final-round election be $(C, V)$, and
    we know in this ``otherwise'' case that $p$ does not uniquely win there.
    
    It is not hard to see that approval is also immune to $\dcpctpnuw$: 
    Let $p$ be a winner of an
    election $(C, V)$. Let $k$ denote the number of ballots (votes) that approve
    of $p$ in $V$. It follows that no candidate is approved by more than
    $k$ votes (else $p$ would not be a winner). Since the \tp\ handling rule
    is used, $p$ can never be eliminated from a subelection as no candidate
    is approved by more votes than $p$. Thus $p$ always proceeds to the final round, and is a winner. So the $\approvaldcpctpnuw$ case follows by
    the above proof with each instance of the word ``unique'' removed and each \te\ and \uw\ 
    respectively changed to \tp\ and \nuw\@.
    
    Since approval is immune to both
    $\ccpctpuw$~\cite{hem-hem-rot:j:destructive-control}
    and $\ccpctpnuw$~\cite{car-cha-hem-nar-tal-wel:j:sct},
    the two constructive cases are analogous, of course by asking respectively as to the 
    $\approvalccpctpuw$ and $\approvalccpctpnuw$ cases whether $p$ is \emph{not}
    a unique winner of $(C, V)$ or not a winner of $(C, V)$, and proceeding in the obvious way,
    again using $(\emptyset, C)$ as our output partition in those cases that do not output
    $\perp$.~\end{proofs}

The following theorem is for the most part due to the fact that certain existing~\cite{mau-rot:j:control-veto-plurality,hem-hem-rot:j:destructive-control} polynomial-time claims for decision problems are proven via algorithms that are already solving the related search 
problem that we care about here.

\begin{theorem}\label{th:poly-time-computable-direct}
The search problem for each of the following control problems is polynomial-time computable:
\begin{enumerate}
    \item $\vetodcpvteuw$.
    \item $\approvaldcpvteuw$.
    \item $\approvalccrpcteuw$.
    \item $\approvalccrpctenuw$.
\end{enumerate}
\end{theorem}
\begin{proofs}
\begin{enumerate}
    \item Maushagen and Rothe~\cite{mau-rot:j:control-veto-plurality} show that
    $\vetodcpvteuw \in \p$ and their algorithm detects whether a solution exists and
    explicitly constructs a solution in polynomial time when a solution exists.

    \item Hemaspaandra, Hemaspaandra, and Rothe~\cite{hem-hem-rot:j:destructive-control}
    show that $\approvaldcpvteuw \in \p$ and their construction detects whether a solution exists and explicitly constructs a solution in polynomial time when a solution exists.

    \item Hemaspaandra, Hemaspaandra, and Rothe~\cite{hem-hem-rot:j:destructive-control}
    show that $\approvalccrpcteuw \in \p$ and their construction detects whether a solution exists and explicitly constructs a solution in polynomial time when a solution exists.

    \item
    The algorithm of Hemaspaandra, Hemaspaandra, and Rothe~\cite{hem-hem-rot:j:destructive-control} drawn on in the previous part is for the UW model and, not surprisingly, does not solve the NUW
    case. We now give a 
    modification of that algorithm,
    and our modified version 
    in fact does, for the NUW case this 
    part is about, construct a solution in 
    polynomial time when a solution 
    exists. Our algorithm 
    proceeds as follows: 
    On input $(C, V, p)$, for each $a \in C$, let $y_a$ be the number of votes in $V$ that 
    approve $a$ and let $Y = \max\{y_a \condition a \in C\}$. 
    If $y_p \neq Y$ and $\| \{a \in C\condition y_a = Y\}\| = 1$, then output $\perp$.
    Otherwise, output $(\{p\}, C-\{p\})$.  Why does $(\{p\}, C-\{p\})$ work?
    Note that in this ``otherwise'' case, we have that either 
   (a)~$y_p = Y$ or (b)~$\| \{a \in C\condition y_a = Y\}\| \geq 2$.  
   If (a)~holds, then the partition $(\{p\}, C-\{p\})$  makes $p$ a winner in the final round, since $p$ will uniquely win and move forward from its subelection, and no candidate has more approvals than $p$ so even if a candidate moves forward from the other subelection it cannot prevent $p$
   from being a final-round winner.
   If~(a)~fails and~(b)~holds, then
   the partition $(\{p\}, C-\{p\})$ will make $p$ a winner (indeed, a unique winner) in the final round, since 
   in the other subelection  there will be at least two candidates that
are approved by $Y$ votes, so they tie as winners of that subelection and are eliminated;
    thus no candidates will move forward from that first-round election.\pagebreak[1] 
    
    The algorithm just given clearly runs in polynomial time.\nopagebreak\end{enumerate}\nopagebreak
    \end{proofs}
We now turn our attention to proving SAT-equivalence.  We start with the following helper theorem.

\begin{theorem}\label{th:sat-equivalent}
Given an election system $\cale$ satisfying $W_\cale \in \p$,
if $\calt$ is one of our partition-control types involving $\cale$ and (the decision problem)
$\calt$ is $\np$-complete, then the search problem for $\calt$ is $\sat$-equivalent.
\end{theorem}
\begin{proofs}
	Assume $\calt$ is one of our partition control types, that $\cale$ is the election system of
    $\calt$, that $W_\cale \in \p$, and that (the decision problem) $\calt$ is $\np$-complete.
	
    $\np$-hard:
    Let $f$ be any refinement of $\widehat{\Pi}_{\calt}$. 
    Since $\calt$ is $\np$-complete, it suffices to show
    $\calt \in \p^f$ to show that $\np \subseteq \p^f$.
    Our polynomial-time algorithm to decide $\calt$ with function oracle $f$
    proceeds as follows: If $x$ is not a triple of the form
    $(C, V, p)$ where $C$ is a set of candidates, $V$ is a vote 
    collection
    over $C$ of the 
    vote-type of $\cale$, and $p\in C$, then reject. Otherwise,
    query the function oracle with $x$ and verify, in polynomial time (using the fact that $W_\cale \in \p$), whether the
    oracle response is a solution to the control problem. If it is, accept, and otherwise reject.

    $\np$-easy:
    It is easy to see that there is a refinement of $\widehat{\Pi}_\calt$, $h$, such that
    $h \in \fp^\np$.
    Let us focus on $\calt$ being a voter partition type.  The candidate cases are exactly 
    analogous except we build the candidate rather than the voter partition.
    On input $(C, V, p)$, $h$ makes sure that the input is of the form $(C, V, p)$, that
    $p \in C$, and that the votes in $V$ are of the type appropriate for $\cale$; if not output
    $\perp$.
    Otherwise, use a single call to $\sat$ to determine if $(C, V, p) \in \calt$. If not, output
    $\perp$. Otherwise, by an easy binary search, with an $\np$ oracle, we will construct a set
    $V_1$ such that $(V_1, V - V_1)$ is a solution to $\calt$ for $(C, V,p)$. In fact, if we 
    naturally encode partitions into binary strings, we can binary search, with any
    $\np$-complete set such as $\sat$ as our oracle, to find the lexicographically least encoding
    of a $V_1 \subseteq V$ such that $(V_1, V - V_1)$ is a solution to $(C, V, p)$ with respect
    to $\calt$, and then we will output $(V_1, V-V_1)$. (The ``helper'' $\np$ set for the binary
    search is simply
    $\{(C, V, p, \calc) \condition (\exists \calc')[\calc' \geq_{\rm lex} \calc$ and
    $\calc'$ encodes a $V_1$ such that $(V_1, V-V_1)$ is a solution to $(C, V, p)$ with respect
    to $\calt]\}$. Since this is an $\np$ set, questions to it can be polynomial-time
    transformed into questions to $\sat$\@.)~\end{proofs}

Theorem~\ref{th:sat-equivalent} is a 
tool that will
help us determine
the search complexity of many of our problems of interest, since
many of those decision problems have been shown to be $\np$-complete in the
literature. Unfortunately, much research on electoral control types has been
in the unique-winner model, and so we need to establish the following new decision-complexity result
before proceeding.

\begin{proposition}\label{p:np-complete}
    As a decision problem,
    $\pluralitydcpctpnuw$ (and equivalently, $\pluralitydcrpctpnuw$)
    is $\np$-complete.
\end{proposition}
\begin{proofs}
    The analogous result in the unique-winner model was established 
    by Hemaspaandra, Hemaspaandra, and Rothe~\cite{hem-hem-rot:j:destructive-control}. Our proof, which we include for completeness, closely follows their proof. We use the same construction (i.e., reduction), but our correctness argument involves 
    small yet important 
    modifications.
    
    Membership in $\np$ is immediately clear.  We 
    now prove $\np$-hardness.
    We in particular provide a reduction from the Hitting Set problem, a known
    $\np$-complete problem~\cite{gar-joh:b:int}.
    
    The Hitting Set problem is defined as follows. Given a set 
    $B = \{b_1, b_2, \ldots, b_m\}$, a family $S = \{S_1, S_2, \ldots, S_n\}$
    of subsets of $B$, and a positive integer $k$, 
    does $S$ have a hitting set of size at most $k$? (That is, 
    is there a set $B' \subseteq B$ with $\|B'\| \leq k$ such that, for each $i$,
    $S_i \cap B' \neq \emptyset$?)
    
    We now state the construction that~\cite{hem-hem-rot:j:destructive-control} used
    for their NP-hardness reduction for Plurality-DC-PC-TP-UW, since we will use the 
    same construction as our NP-hardness reduction
    for Plurality-DC-PC-TP-NUW\@.
    \begin{construction}[\cite{hem-hem-rot:j:destructive-control}]
    
    Given a triple $(B, S, k)$, where $B=\{b_1, b_2, \ldots, b_m\}$, 
    $S = \{S_1, S_2, \ldots, S_n\}$ is a family of subsets of $B$, and $k \leq m$
    is a positive integer, construct the following election:
    
    \begin{enumerate}
        \item The candidate set is $C = B\cup \{c, w\}$.
        \item The vote set $V$ is defined as:
        \begin{enumerate}
            \item There are $2(m-k) + 2n(k+1) + 4$ votes of the form
            $c>w>\cdots$, where the ``$\cdots$'' means that the remaining candidates are in some arbitrary order.
            \item There are $2n(k+1)+5$ votes of the form $w > c> \cdots$.
            \item For each $i \in  \{1, \ldots, n\}$, there are $2(k+1)$
            votes of the form $S_i > c > \cdots$, where ``$S_i$'' denotes
            the elements of $S_i$ in some arbitrary order.
            \item For each $j \in \{1, \ldots, m\}$, there are two
            votes of the form $b_j > w > \cdots$.
        \end{enumerate}
        \item The distinguished candidate is $c$.
    \end{enumerate}
    
    \end{construction}
    
    We now state two claims that will be used to prove
    that the reduction works in our case.

    \begin{claim}[\cite{hem-hem-rot:j:destructive-control}]\label{claim:forward}
    If $B'$ is a hitting set of $S$ of size $k$, then $w$
    is the unique winner of the plurality election $(B' \cup \{c, w\}, V)$.
    \end{claim}
    
    \begin{claim}\label{claim:backward}
        Let $D \subseteq B \cup \{w\}$. If $c$ is not a winner of plurality
        election $(D\cup\{c\}, V)$, then there exists a set $B'\subseteq B$
        such that 
        \begin{enumerate}
            \item $D = B' \cup \{w\}$,
            \item $w$ is the unique winner of plurality election
            $(B' \cup \{c, w\}, V)$, and 
            \item $B'$ is a hitting set of 
            $S$ of size less than or equal
            to $k$.
        \end{enumerate}
    \end{claim}
    \begin{proofsof}{of Claim~\ref{claim:backward}}
        Fix $D \subseteq B\cup\{w\}$ such that $c$ is not a winner 
        of plurality election $(D\cup\{c\}, V)$. We will show that the above
        three properties hold, by using a modified version of the argument
        used in~\cite{hem-hem-rot:j:destructive-control}.
        
        For a given election, we will let $\score(d)$ denote the number of
        votes that rank candidate $d$ first in that election.
        
        First, notice that for each $b \in D\cap B$, $\score(b) < score(c)$
        in $(D\cup \{c\}, V)$. Since $c$ is not a winner of that election, 
        it must hold that $w$ is the unique winner of that election, and thus
        $\score(w) > \score(c)$.
        
        Let $B' \subseteq B$ be such that $D = B' \cup \{w\}$. Then
        $D\cup\{c\} = B' \cup \{c, w\}$. Thus it follows that $w$
        is the unique winner of the election $(B'\cup \{c, w\})$, proving the first
        two properties. 
        
        Finally, observe that in $(B'\cup\{c, w\}, V)$, it holds that
        \begin{enumerate}
            \item $\score(w) = 2n(k+1) + 5 + 2(m-\|B'\|)$, and
            \item $\score(c) = 2(m-k) + 2n(k+1) + 4 + 2(k+1)\ell$,
        \end{enumerate}
        where $\ell$ is the number of sets in $S$ that 
        have an empty
        intersection with $B'$, i.e., are not ``hit by $B'$.''
        Since $w$ is the unique winner of the election, it follows that
        \begin{align*}
            \score(c) &< \score(w),\\
            2(m-k) + 2(k+1)\ell &< 1 + 2(m-\|B'\|),\mbox{ and}\\
            (k+1)\ell + \|B'\| - k &< 1/2.
        \end{align*}
        Since $\ell$ is a nonnegative integer, the only value that it can have here is 0, and so it follows that 
        $B'$ is a hitting set of $S$ of size at most $k$.~\end{proofsof}

    Now to conclude the proof of Proposition~\ref{p:np-complete}, we will leverage the two claims above
    to show that the following two items are equivalent: 
    \begin{enumerate}
    \item There is a set $B'\subseteq B$ of size at most $k$ that is a hitting
    set of $S$.
    \item There is a partition of $C$ such that
    $c$ can be prevented from being a winner of the two-stage plurality election conducted 
    under the PC-TP-NUW model.
    \end{enumerate}
    
    Let $B'$ be a hitting set of $S$ of size $k$. Let $C_1 = B' \cup \{c, w\}$ 
    and $C_2 = C - C_1$. By Claim~\ref{claim:forward} it holds that $w$ is the unique winner
    of the subelection $(C_1, V)$, and thus $c$ does not proceed to the final round and is not a winner.
    
    Suppose there is a partition of $C$ such that $c$ is not a winner of the
    corresponding two-stage plurality election. 
    It must hold that $c$ 
    is eliminated 
    either in the first-round 
    election or in the final-round election.
    Then it holds that there is a set $D \subseteq B \cup \{w\}$ such 
    that $c$ is not a winner of plurality election $(D\cup\{c\}, V)$. It 
    directly follows from Claim~\ref{claim:backward} that 
    $S$ has a hitting set of size at most $k$. This concludes
    the proof that $\pluralitydcpctpnuw$ is $\np$-complete.
    
    The ``equivalently'' follows from the fact that for every election system $\cale$, $\caledash\allowbreak\dcrpctpnuw=\caledash\allowbreak\dcpctpnuw$~\cite{hem-hem-men:j:search-versus-decision}.~\end{proofs}

\begin{corollary}\label{cor:sat-equivalent}
	The 
	search-problem versions of the 
	following 
	are  $\sat$-equivalent.
	\begin{enumerate}
		\item %
  $\pluralitydcpctpnuw$
		\item $\pluralitydcrpcteuw$
		\item $\vetodcrpctpnuw$
		\item $\vetodcrpctenuw$
	\end{enumerate}
\end{corollary}
\begin{proofs}

Each of these four is NP-complete, the first by 
Proposition~\ref{p:np-complete}, the second is from~\cite{hem-hem-rot:j:destructive-control}, and the remaining two are from Maushagen and Rothe~\cite{mau-rot:j:control-veto-plurality}.
	The result follows from those four NP-completenesses,
	by Theorem~\ref{th:sat-equivalent}.~\end{proofs}
Table~\ref{table:summary} provides, 
for the concrete election systems discussed in this paper,
a summary of our results on the search complexity of each 
equivalence class of decision-collapsing
control problems. 
This paper's results 
establish that within each such equivalence class the search
complexities must be 
polynomially related; the table is summarizing 
our results on whether 
those linked complexities are both ``polynomial(-time computability)'' 
or are both ``SAT-equivalence.'' %
(The leftmost column's classification of which control types, for the three 
studied election systems, collapse as decision problems is due to
Carleton et al.~\cite{car-cha-hem-nar-tal-wel:j:sct}, building on
Hemaspaandra, Hemaspaandra, 
and Menton~\cite{hem-hem-men:j:search-versus-decision}.)

\newcommand{\rowonewidth}{60mm}
\newcommand{\rowtwowidth}{26mm}
\newcommand{\rowthreewidth}{51mm}

\newcommand{\mysizedrow}[4]
{\parbox[c][#1][c]{\rowonewidth}{\raggedright#2} & 
\parbox[c][#1][c]{\rowtwowidth}{#3} &
\parbox[c][#1][c]{\rowthreewidth}{#4}}

\begin{table}[p]%
\centering
\rowcolors{1}{white}{lightgray}

\begin{tabular}{p{\rowonewidth}|p{\rowtwowidth}|p{\rowthreewidth}}
\mysizedrow{1.5cm}{\centering Collapsing Decision Control Types}{%
\centering
Complexity of Their Search Versions}{\centering References} \\\hline
\mysizedrow{1cm}{$\pluralitydcrpctpnuw$, $\pluralitydcpctpnuw$}{$\sat$-equivalent} 
{Corollaries~\ref{cor:concrete-relationships-from-hhm} and~\ref{cor:sat-equivalent}}\\
\mysizedrow{19mm}{$\pluralitydcrpctenuw$, $\pluralitydcrpcteuw$, $\pluralitydcpctenuw$, $\pluralitydcpcteuw$}
{$\sat$-equivalent}{Corollaries~\ref{cor:concrete-relationships-from-hhm} and~\ref{cor:sat-equivalent}}\\
\mysizedrow{1cm}{
$\vetodcpvtenuw$,\linebreak$\vetodcpvteuw$}{polynomial}{Corollary~\ref{cor:relationships-from-priors} and Theorem~\ref{th:poly-time-computable-direct}}\\
\mysizedrow{1cm}{$\vetodcrpctpnuw$,\linebreak$\vetodcpctpnuw$}{$\sat$-equivalent} 
{Corollaries~\ref{cor:concrete-relationships-from-hhm} and~\ref{cor:sat-equivalent}}\\
\mysizedrow{15mm}{$\vetodcrpctenuw$, $\vetodcrpcteuw$, $\vetodcpctenuw$, $\vetodcpcteuw$}{$\sat$-equivalent}{Corollaries~\ref{cor:concrete-relationships-from-hhm} and~\ref{cor:sat-equivalent}}\\
\mysizedrow{1cm}{$\approvaldcrpctpnuw$, $\approvaldcpctpnuw$}{polynomial} 
{Corollary~\ref{cor:concrete-relationships-from-hhm} and 
Theorem~\ref{th:approval-immune-poly-time}}\\
\mysizedrow{29mm}{$\approvaldcrpctpuw$, $\approvaldcpctpuw$, $\approvaldcrpctenuw$, $\approvaldcrpcteuw$, $\approvaldcpctenuw$, $\approvaldcpcteuw$}
{polynomial}{Corollary~\ref{cor:from-uwarp-to-approval}  and 
Theorem~\ref{th:approval-immune-poly-time}} \\
\mysizedrow{1cm}{$\approvaldcpvteuw$, $\approvaldcpvtenuw$}{polynomial} {Corollary~\ref{cor:relationships-from-priors} and Theorem~\ref{th:poly-time-computable-direct}}\\
\mysizedrow{1cm}
{$\approvalccrpctpuw$, $\approvalccpctpuw$}{polynomial}{Corollary~\ref{cor:from-uwarp-to-approval} and 
Theorem~\ref{th:approval-immune-poly-time}}\\
\mysizedrow{1cm}{$\approvalccrpctpnuw$, $\approvalccpctpnuw$}{polynomial} {Corollary~\ref{cor:relationships-from-priors} 
and Theorem~\ref{th:approval-immune-poly-time}}\\
\mysizedrow{1cm}
{$\approvalccrpcteuw$, $\approvalccpcteuw$}{polynomial}{Corollary~\ref{cor:relationships-from-priors} and 
Theorem~\ref{th:poly-time-computable-direct}}\\
\mysizedrow{1cm}{$\approvalccrpctenuw$, $\approvalccpctenuw$}{polynomial} {Corollary~\ref{cor:relationships-from-priors} 
and Theorem~\ref{th:poly-time-computable-direct}}\\\hline
    \end{tabular}
    \caption{For each collection of (decision-problem) collapsing partition control types for 
    plurality, veto, and approval elections, we have shown that their search problems are also of 
    the same complexity. This table, for each, states whether the search problems are 
    polynomial-time solvable (denoted ``polynomial'' in the table) or are $\sat$-equivalent.
    Proposition~\ref{p:sat-equivalent} and Proposition~\ref{p:polynomial} are not listed
    in the References column, since it is clear when they
    are being drawn on (see the discussion of this near the 
    end of Section~\protect\ref{s:results}).} \label{table:summary}
\end{table}

\afterpage{\clearpage}

Note that Table~\ref{table:summary} gives various results that 
are not explicitly stated in any of the theorems of this paper, but that in fact follow from 
combining results in the 
paper.  To help explain why those results are indeed
validly included in the table, we first need to state 
two results, and then will come back to how they
help populate the table.

While the following two results may seem intuitive, for the sake of completeness we provide their proofs, since we are quietly using them in populating the table.

\begin{proposition}\label{p:sat-equivalent}
    Let 
    search problems
    $A$ and $B$ be 
    polynomially search-equivalent.
    If $A$ is \sat-equivalent, then $B$ is \sat-equivalent.
\end{proposition}
\begin{proofs}
    Fix two polynomially search-equivalent 
    search
    problems $A$ and $B$, with
    $A$ being \sat-equivalent. We need to show that, per our definitions,
    $B$ is both \np-hard and \np-easy. 
     $\np$-hard:
     Let $g$ be an arbitrary refinement of $\widehat{\Pi}_B$. We
    need to show that $\np \subseteq \p^g$. It suffices to show that $\sat \in \p^g$. 

    Let $r$ be a function that witnesses $A \polysearchreduce B$. It follows
    that the function
    $h$ that on arbitrary input $x$ maps to $\perp$ if $g(x)$ maps to $\perp$,
    and that otherwise maps to $r(x, g(x))$, is a 
        refinement of $\widehat{\Pi}_A$.
        
    Since $A$ is NP-hard, we have that 
     $\sat \in \p^h$.  Let $T$ denote the 
     polynomial-time algorithm (with oracle $h$)
     that decides $\sat$.
     Here is the $\p^g$ algorithm $T'$ for $\sat$. 
     Our polynomial-time algorithm $T'$ (with oracle $g$)
     simulates $T$, and each time $T$ asks a 
     question, $y$, to its oracle, $T'$ asks the same 
     question to its oracle $g$, and if the answer 
     is $\perp$ then $T'$ acts as if the answer $T$
     got is $\perp$, and 
     otherwise $T'$ acts as if the answer $T$ got 
     is $h(y)$.  So the outcome of $T'$ with oracle $g$ is precisely the same
     as that of $T$ with oracle $h$, and so we have 
     proven
     that 
     $\sat \in \p^g$.

    \np-easy: It suffices to show that $\widehat{\Pi}_B$ has a refinement in $\fp^\sat$. 
    Let $s$ denote a function witnessing $B \polysearchreduce A$. Since
    $A$ is $\sat$-equivalent, there is a refinement $t$ of $\widehat{\Pi}_A$
    such that $t \in \fp^\sat$. But it also holds, for each $x$, that
    $t(x)$ is a solution for $x$ under $A$ if and only if %
    ($t$ is not $\perp$ and) $s(x,t(x))$ is a solution
    for $x$ with respect to $B$. 
    Thus the function, $h$, that on
    input $x$ is $\perp$ if $t(x)$ is $\perp$
    and otherwise is $s(x,t(x))$ 
    is a refinement
    of $\widehat{\Pi}_B$. But since $t \in \fp^\sat$,
    clearly so also is $h$.~\end{proofs}
\begin{proposition}\label{p:polynomial}
    Let 
    search problems 
    $A$ and $B$ be 
    polynomially search-equivalent.
    If $A$ is polynomial-time computable, then $B$ is 
    polynomial-time computable.
\end{proposition}
\begin{proofs}
    Fix two polynomially search-equivalent 
    search
    problems $A$ and $B$, with
    $A$ being polynomial-time computable. We need to show that
    $B$ is polynomial-time computable, by providing a refinement of
    $\widehat{\Pi}_B$ in $\fp$. 

    Let $f$ be a polynomial-time computable refinement of $\widehat{\Pi}_A$
    and let $g$ be a function witnessing $B \polysearchreduce A$. 
    The function $h$ that on input $x$ is $\perp$ if 
    $f(x)$ is $\perp$ and otherwise is $g(x,f(x))$
    is clearly a polynomial-time computable refinement of $\widehat{\Pi}_B$.~\end{proofs}

As mentioned earlier, Table~\ref{table:summary} gives various results that 
are not explicitly stated in any of the theorems of this paper, but that in fact follow from 
combining results in the 
paper.  For example, 
consider Plurality-DC-RPC-TE-NUW\@.
The table states that it is 
SAT-equivalent, and that 
is true by the reasoning 
implicit in the table's right-hand column.
In particular, by 
Corollary~\ref{cor:sat-equivalent} Plurality-DC-RPC-TE-UW
is SAT-equivalent, and so 
by Corollary~\ref{cor:concrete-relationships-from-hhm}
and Proposition~\ref{p:sat-equivalent} it follows that
Plurality-DC-RPC-TE-NUW
is SAT-equivalent.  
In fact, 
Proposition~\ref{p:sat-equivalent} is broadly used 
in the table as to problems stated to be SAT-equivalent,
namely to ``inherit'' the SAT-equivalence from one 
problem to other problems that we have shown 
to be search-equivalent 
to it.
Likewise, Proposition~\ref{p:polynomial} is analogously used in that table, regarding many cases of polynomial-time 
computable search problems.

Section~\ref{s:results} has now achieved the 
broad set of results that its first paragraph promised, and as will also be summarized in the Conclusion 
section.

\section{Conclusion}
In this paper, for all collapsing electoral control types found in 
Hemaspaandra, Hemaspaandra, and Menton~\cite{hem-hem-men:j:search-versus-decision} and
Carleton et al.~\cite{car-cha-hem-nar-tal-wel:j:sct}---those are all such collapses for the domains
and groupings studied there (see~\cite{hem-hem-men:j:search-versus-decision} and 
especially~\cite{car-cha-hem-nar-tal-wel:j:sct})---we proved that even the 
\emph{search-problem}
complexities of the collapsing types are polynomially equivalent (given oracle access to the election system's winner problem).

In doing this---and building on earlier notions of relating search problems---we defined
reductions that, for the case of collapsing electoral control problems,
express how solutions to one can be efficiently converted to solutions to the other. 

Also, for the key concrete systems plurality, veto, and approval, for each of their collapsing pairs 
we establish either that both of the pair's search versions are polynomial-time computable, or that 
both of the pair's search versions
are $\sat$-equivalent (see Table~\ref{table:summary}).
An interesting open direction would be, for that part of our work, to seek more general results, such as dichotomy theorems
covering broad collections of election systems. However, that may be difficult since not much is known as to dichotomy
theorems even for the decision cases of (unweighted) control problems 
(however, see~\cite{hem-hem-sch:c:psrs,hem-sch:c:psrs,fal-hem-hem:j:weighted-control}), though the few known such cases would be natural
starting points to look at in this regard.

\subsection*{Acknowledgments} 
Preliminary versions of this work appeared in 
\emph{Proceedings of the 22nd International Conference on Autonomous Agents and Multiagent Systems}~\cite{car-cha-hem-nar-tal-wel:c-aamas:search-vs-search} and in 
\emph{Proceedings of the 21st European Conference on Multi-Agent Systems}~\cite{car-cha-hem-nar-tal-wel:c-eumas:search-vs-search}. We thank the anonymous referees for their helpful comments.
This work is supported in part
    by NSF grants CCF-2006496
and DUE-2135431,
    CIFellows grant CIF2020-UR-36, and a Renewed Research Stay grant from the Alexander von Humboldt Foundation.

\bibliographystyle{alpha}
%
%
%
%
\newcommand{\etalchar}[1]{$^{#1}$}

\end{document}